\documentclass[fleqn,usenatbib]{mnras}

\usepackage{newtxtext,newtxmath}
\usepackage[T1]{fontenc}

\DeclareRobustCommand{\VAN}[3]{#2}
\let\VANthebibliography\thebibliography
\def\thebibliography{\DeclareRobustCommand{\VAN}[3]{##3}\VANthebibliography}

\usepackage{graphicx}	% Including figure files
\usepackage{amsmath}	% Advanced maths commands
\DeclareMathOperator*{\argmin}{arg\,min}

\usepackage{xcolor}
\usepackage{booktabs}
\usepackage{placeins}

\newcommand{\fnl}{f_{\rm NL}}
\newcommand{\EBV}{E(B\!-\!V)}
\newcommand{\sigclean}{\sigma_{\rm clean}}
\newcommand{\etaadd}{\eta_{\rm add}}

\newcommand{\alttext}[1]{\sbox0{\par\smallskip\noindent{\small Alt text: #1}}}

\title[Deprojection of Angular Imaging Systematics]{Information-Criterion-Driven, Non-linear Deprojection of
Angular Imaging Systematics for the SPHEREx Galaxy Clustering Sample}

\author[J. Cheshire et al.]{
James R. Cheshire IV,$^{1}$\thanks{E-mail: cheshire@caltech.edu}
Sean T. Bruton,$^{1}$
and Charuhas W. Shiveshwarkar$^{2,3}$
\\
$^{1}$Department of Physics, California Institute of Technology, 1200 E. California Blvd, Pasadena, CA 91125, USA\\
$^{2}$Center for Cosmology and Astroparticle Physics, The Ohio State University, 191 West Woodruff Avenue, Columbus, OH 43210, USA\\
$^{3}$Department of Physics, The Ohio State University, 191 West Woodruff Avenue, Columbus, OH 43210, USA
}

\date{Accepted XXX. Received YYY; in original form ZZZ}

\pubyear{\the\year{}}

\begin{document}
\label{firstpage}
\pagerange{\pageref{firstpage}--\pageref{lastpage}}
\maketitle

\begin{abstract}
Imaging systematics that inject false, or suppress true, galaxy clustering signals are of primary
concern for current- and next-generation galaxy surveys, such as
the Spectro-Photometer for the History of the Universe, Epoch of Reionization, and Ices Explorer
(SPHEREx) in its search for local primordial non-Gaussianity $\fnl$. As
statistical errors shrink with growing galaxy samples, systematic rather than statistical uncertainties come to
dominate the cosmological error budget. For effects whose
spatial patterns are known or well-modelled \emph{a priori}, a common mitigation approach regresses
(deprojects) the density variation associated with a known angular template. In this work we present
and demonstrate novel aspects of this procedure as will be applied to the SPHEREx galaxy clustering
sample. These include flexible -- but, crucially, constrained and interpretable -- non-linear basis
regression with polynomial/spline bases, and aggressive information criterion-driven
automated template selection with an associated unbiased two-stage regression, using agreement
between two independent selectors as a consistency diagnostic. The method automatically selects a
template set of minimal-but-sufficient complexity while limiting the number of templates actually
regressed, minimising cosmological information loss and making modern additive-plus-multiplicative
approaches tractable. We demonstrate these features on SPHEREx-like lognormal mock catalogues
with injected contamination of various spatial profiles, amplitudes, and levels
of complexity. Principled, signal-preserving deprojection of the kind
demonstrated here is a crucial step towards achieving SPHEREx's goal of $\sigma(\fnl) < 0.5$.
\end{abstract}

% Select between one and six approved MNRAS keywords.

%%%%%%%%%%%%%%%%%%%%%%%%%%%%%%%%%%%%%%%%%%%%%%%%%%

%%%%%%%%%%%%%%%%% BODY OF PAPER %%%%%%%%%%%%%%%%%%

\section{Introduction}
\label{sec:intro}

Current- and next-generation large-scale structure (LSS) surveys, such as DESI~\citep{adame2025},
\emph{Euclid}~\citep{euclid2025}, Rubin/LSST~\citep{ivezic2019}, Roman~\citep{eifler2021}, and SPHEREx~\citep{dore2014,bock2026} are or
soon will be revolutionising our understanding of the assembly of matter in the
Universe with unprecedented statistical samples. With each increase in
statistical fidelity, the onus of controlling sources of systematic uncertainty
is elevated proportionally, and these unprecedented surveys therefore demand
unprecedented levels of care in this regard. Large angular scales
(small wavenumber $k$) are where a large amount of cosmological information
resides, particularly the scale-dependent bias signature of local primordial
non-Gaussianity, $\propto \fnl/k^{2}$~\citep{dalal2008,dore2014}, which is of particular
interest to SPHEREx; however, these same large angular scales are those most
severely afflicted by the typical imaging systematics of large-scale structure
surveys.

A forced-photometry experiment, as is represented by SPHEREx's nominal
``Level 4'' cosmology mission~\citep{bock2026}, may inherit effects on its galaxy
selection from the reference catalogue parent surveys that define its target list
(which may have sharp or discontinuous features at survey boundaries), as well
as from typical astrophysical effects such as dust extinction, stellar
contamination, and zodiacal emission, which generally vary smoothly on the sky, and SPHEREx's own
spectrophotometric instrumental effects and selection properties. The addition
of misidentified stellar contaminants in a galaxy catalogue, for instance, or even the
nigh-unavoidable effect of galaxy sample density loss in crowded stellar regions
due to blends and contamination from bright stellar neighbors, will produce a
strong variation of power on the large angular scales where the cosmological
signal of interest is also strongest.

The current most stringent, but increasingly cosmic variance-limited, constraints on
$\fnl$ are $-0.9 \pm 5.1$, obtained from the cosmic microwave background
bispectrum~\citep{planck2018ng}. Surpassing this level with data from LSS surveys requires
extreme care in approaching contamination, with SPHEREx requiring residual
clustering systematics to be controlled to a level of $\delta n / \bar{n} < 0.2$
per cent rms per dex in comoving scale~\citep{dore2014} to reach its target of
$\sigma(\fnl) < 0.5$.

Various approaches to clustering systematic mitigation have been explored and
employed in the past. A common technique is to simply cut the angular scales
that are suspected of being contaminated (\emph{e.g.}, setting a strict $k_{\rm min}$)~\citep{agarwal2014,ho2015},
which is blind to the nature of the systematic though comes at the significant
cost of signal loss. More sophisticated
approaches along this line seek to use information about the geometry of the
survey/systematics to make more ``surgical'' sets of cuts~\citep{hand2017,zhao2021,wen2025,bruton2026}. Another approach is to forward model the signal transfer function/selection function
empirically using synthetic injected sources~\citep{everett2022,kong2025}, a
technique which provides extremely high-fidelity measurements of certain effects, though is insensitive
to others (astrophysical or reference catalogue-induced effects, for instance) and
comes with significant computational cost.

This paper focuses on a third common approach, which relies upon
contaminating effects with well-known/well-measured spatial patterns. For a
contaminant whose pattern on the sky and spectral properties are known \emph{a priori},
such as dust extinction or, even more cleanly, a parent survey depth pattern,
one may construct a spatial \emph{template} of the contaminating effect, and
attempt to account for the modulation of the measured density field due to this
effect directly, based on some model for how the contaminant couples to the
measured field. These ``template deprojection'' approaches typically vary in terms of the array and type
of templates used, the differing assumptions in how the model couples/is fit,
and the approaches taken to mitigate or characterize cosmological signal loss.

The earliest widely used approaches along this line iterated over individual
systematic template maps, assigning each galaxy a weight based on the relative
contribution in that template relative to the mean density, assuming a linear
coupling between the systematic and the density field~\citep{ross2011,ross2012}. However, this
assumes perfect separability and low correlation between templates (an
assumption that is strongly violated by multiple effects which are themselves
all correlated with the structure of the Galactic plane). Mode projection seeks
to marginalize analytically over the amplitude of the template at a 2-point
level~\citep{rybicki1992,slosar2004,leistedt2013,leistedt2014,elsner2016}, which is equivalent to a covariance-aware deprojection, and is
still linear in the template. In their default configurations, these approaches
treat every template equally as worthy of inclusion/regression, which tends to have
the unintended effect of increasing signal loss directly in proportion with the
number of additional templates~\citep{elsner2016,elsner2017}. For the increasing number of effects and
potential templates considered by modern surveys, this represents an
increasingly tangible practical barrier.

One approach to overcoming this hurdle is to grant additional degrees of fitting
freedom to the density-template relation by relaxing the assumptions of linear
coupling. This is typically done with neural network or gradient-boosted
tree methods~\citep{rezaie2020,rezaie2021,chaussidon2022} which attempt to learn arbitrary non-linear mappings between
the template vector and the expected density. Another approach is to sparsify
the template vector itself using regularized regression approaches.
\citet{wh21} (hereafter WH21) introduce this approach, employing and comparing the ElasticNet
(EN)~\citep{zou2005} fitting approach with an iterative forward-selection (IFS)-based
algorithm, demonstrating that both methods show substantially reduced signal
loss relative to deprojecting the full array.

The non-linear and sparse approaches each address particular issues with the overall template
deprojection method, but are in mild tension with one another. The aggressive
non-linear approaches exhibit (not accidentally) extreme flexibility in what
they are able to remove, up to and including significant amounts of cosmological
signal that one would otherwise like to preserve. On the other hand, the
regularized-linear approaches will be more conservative in terms of signal
preservation, at the cost of potentially not allowing flexibility to capture
sharp or complicated non-linear relationships in the template-density field
coupling. One might therefore endeavor to devise a procedure that captures the
most favorable aspects of each approach, where the deprojection has the freedom
to be moderately non-linear where the data demand it, is linear otherwise, and is sparse
throughout.

In this work, we develop and validate such a method in the context of the upcoming SPHEREx galaxy
clustering sample. The novel features of this method are the following:

\begin{enumerate}
\item \textbf{Information-criterion-driven selection with unbiased two-stage
        regression.}
        Regularized regression algorithms such as ElasticNet are, in the
        clustering systematics context, being asked to perform two tasks which
        are slightly at-odds with one another. In using a penalty term for
        selection, it also induces a shrinkage effect on the regression
        coefficients for a given template, leading to a suppression of recovered
        systematic signal in simulations and potentially elevated residuals in
        the post-cleaned galaxy field. We instead decouple the two tasks, in a manner similar
        to the post-lasso estimation of \citet{belloni2013}, but with selection scored by a
        calibrated information criterion, rather than the \texttt{scikit-learn} default
        of held-out mean squared error; this \emph{selection} step determines
        the list of templates which genuinely provide additional information to
        the fit as informed by the underlying data vector, with a single scale
        parameter that may be calibrated to a desired false-positive ratio based
        on deprojecting uncontaminated mocks. The second step employs, by default, a standard
        multivariate ordinary least-squares (OLS) using the ensemble of templates
        selected by the first step, which recovers systematic amplitude
        coefficients without the bias that may be incurred by sparsified estimators. We employ a structurally
        different selection algorithm, a variant of the IFS algorithm from WH21, as an
        independent consistency diagnostic. 
\item \textbf{non-linear basis expansion with automatic order determination.}
        Rather than assume a purely linear fitting basis (template-systematic coupling), we decompose
        each template into a Legendre polynomial or B-spline basis
        prior to the selection step. The fit is then non-linear in the template
        values while remaining linear in the fit coefficients, allowing us to
        exploit the body of existing linear regression machinery and preserve
        the interpretability of the chosen set. The non-linear basis expansion and
        the automated selection naturally work in concert, as demonstrated in
        \S\ref{sec:res-non-linear}, feeding-in templates with polynomial coefficients
        up to $K=6$ and allowing the selection machinery to determine the
        necessary polynomial order. The complexity-penalized selection truncates
        each series to a minimal-but-sufficient set, allowing the complexity to
        be set directly by the data on a template-by-template basis, and to be
        read-off directly. A standard
        linear coupling preserves only the first-order term, thereby directly
        reducing to the standard deprojection case. B-spline bases are employed
        in instances where there is an obvious physical motivation to do so,
        such as to place knots at the systematic value corresponding to a sharp
        boundary resulting from a parent imaging survey sample cut.
\item \textbf{Composition with joint additive-multiplicative correction.}
        Recent works (\emph{e.g.},~\citealt{wsrs26}) have emphasized the importance of jointly treating
        additively and multiplicatively coupled systematics. We extend the joint estimator of \citet{bmds24} with a treatment of shot noise
        necessary for realistic SPHEREx pixelisations and source densities. On mocks with $\sim 1\%$ stellar contamination, the
        joint estimator correctly attributes injected additive and multiplicative
        contamination as measured by the second-moment statistic. The automated down-selection
        procedure ensures the number of fit templates is small, making this joint fit practical
        for an expanded template set.
\end{enumerate}

This paper describes the (2-dimensional component of the) clustering systematic
deprojection methodology that will be employed for the SPHEREx galaxy clustering
sample, as applied to and validated on mocks. We do not directly propagate
post-deprojection residuals or uncertainties to $\fnl$ or attempt to measure or
quantify the level of contamination on data, and we do not address the
associated null test and unblinding framework which would be necessary to
do so in a rigorous way; the latter will be the subject of
a dedicated future publication. We also do not describe the estimation of radial
density modulation and spectrophotometric forward-modelled systematic templates,
which will likewise be the subject of their own work (though we note that the
generalisation of projection onto the ElasticNet basis from a set of pixels to
one of voxels is trivial, given an appropriate template with radial structure).

Section~\ref{sec:methods} describes the deprojection framework, automated
selection procedure and calibration, and non-linear basis expansion. Section~\ref{sec:mocks} describes the lognormal mock galaxy
catalogues, the model of injected contamination, and describes the template
taxonomy used for this validation exercise. Section~\ref{sec:results} presents
the main results, examining template deprojection performance, systematic
amplitude recovery, and signal preservation and comparing to published methods.
Finally, we conclude in Section~\ref{sec:conclusions}. HEALPix pixelisation~\citep{healpix} is employed throughout, using the \texttt{NEST} pixel ordering in an
equatorial coordinate system.

\section{Deprojection methodology}
\label{sec:methods}

\subsection{Contamination and correction model}
\label{sec:framework}
We model the modulation of a true underlying galaxy density field $n_g$ in a
pixel $p$ by a
multiplicatively coupled systematic field $m$, with an additional spurious
additive component $A$, as
\begin{equation}\label{eq:nobs}
  n_{\rm obs}(p) = n_g (p) \left[ 1 + m(p) \right] + A(p).
\end{equation}
Many commonly-worried-about systematic effects will couple
multiplicatively, due to a systematic modifying the detection probability of a
given galaxy, which can be driven by a variety of effects such as dust
extinction, local background/confusion, and observing depth. This multiplicative
coupling is a key assumption of many of the historical approaches to template
deprojection, under the assumption that $m(p)$ is relatively small. Additive
effects are the manifestation of mechanisms which directly add or remove objects from
the catalogue, such as misidentified stars, artifacts in the reference photometry
catalogue, and some manifestation of object blends (the latter being a
particularly pernicious effect which couples both multiplicatively and
additively). These effects enter the 2-point function differently, with
multiplicative terms being in proportion to the clustering signal while additive
terms scale as $a^2 C_{\ell}^{tt}$ for a template $t$ (\citealt{bmds24},
hereafter BMDS24). Given
sufficient control of additive systematics (through, for instance, careful
star-galaxy separation), these additive terms are often small and their mean contributions may be absorbed by additional degrees of freedom in the deprojection fit (Section~\ref{sec:res-addmult}), though in general, and particularly for the
sensitivity targets of next-generation experiments, their higher-order effects may become more
difficult to ignore.

\subsubsection{Density contrast}\label{sec:contrast}
The primary observable for LSS surveys is the galaxy density contrast
$\delta_{\rm obs} = n_{\rm obs}/\bar{n}_{\rm obs} - 1$, where the bar denotes a mean over pixels (which in this case may itself be affected by contamination). We may characterize the
additive field in terms of the fraction of the sample that is spurious,
\begin{equation}\label{eq:etaadd}
  \eta_{\rm add} = \bar{A}/\bar{n}_{\rm obs},
\end{equation}
and its spatial fluctuations
$\delta_A \equiv (A - \bar{A})/\bar{n}_{\rm obs}$. A spurious population with
no net mean $\eta_{\rm add}$ may still have spatial structure. We may then write the total change in number density due to both the spurious population and multiplicative contaminants $\delta_m \equiv m - \bar{m}$ as $\bar{n}_g/\bar{n}_{\rm obs} = (1-\etaadd)/(1 + \bar{m} + \overline{\delta_g \delta_m}) \equiv 1 - \eta$, allowing us to write Eq.~\ref{eq:nobs} as
\begin{equation}\label{eq:deltaobs}
  \delta_{\rm obs} = (1 - \eta)(1 + m) \delta_g + (1 - \eta)\delta_m + \delta_A - (1-\eta) \overline{\delta_g \delta_m}.
\end{equation}
Only the spatially varying component of $m$ is observable, with $\bar{m}$ absorbed by $\eta$ in practice, so we take $m$ to be zero-mean; if we likewise assume the multiplicative contaminant field and the galaxy field to be independent such that $\langle\overline{\delta_g \delta_m} \rangle= 0$ (where the angle brackets $\langle \cdot \rangle$ denote the average over galaxy realizations with systematics held fixed), then we may make the additional simplification $1-\eta \approx 1 - \etaadd$, which we will use hereafter. Eq.~\ref{eq:deltaobs} is Eq.~2 of \citet{wsrs26} (hereafter WSRS26) written in
terms of the fields in Eq.~1, identifying $f_{\rm mult} = (1-\etaadd)m$
and $f_{\rm add} = \delta_A$ with their contamination fields. In the case of
$\eta = 0$, this is equivalent to Eq.~2 of BMDS24. The multiplicative field of Eq.~\ref{eq:deltaobs} enters
twice: once modulating the clustering signal, and again as a direct imprint on
the contrast -- a region where galaxies are preferentially lost has depressed
observed density regardless of the underlying structure. The additive field
enters only in this second way, so a correction operating on the first moment of
a density contrast map sees only the sum of the two mechanisms and is unable to
separate them. This requires leveraging higher moments of the map, as described
in Section~\ref{sec:regress}.

We define the first-moment systematic contamination $\delta_{\rm sys} \equiv (1-\etaadd)m + \delta_A$. For $\delta_g$ independent of $\delta_m$ and $\delta_A$, we may obtain the 2-point function by averaging over the galaxy field with the systematics held fixed. Since $\langle \delta_g \rangle = 0$, the cross terms vanish, giving
\begin{equation}\label{eq:2pt}
\begin{aligned}
  \langle \delta_{\rm obs}(p)\,\delta_{\rm obs}(p') \rangle
  = (1&-\eta_{\rm add})^2\,w_g(\theta_{pp'})\,[1+m(p)]\,[1+m(p')]\\
  &+ \delta_{\rm sys}(p)\,\delta_{\rm sys}(p') + \delta_{pp'}\,\frac{1+\delta_{\rm sys}(p)}{\bar{N}_{\rm obs}},
\end{aligned}
\end{equation}
where $\theta_{p p'}$ is the angular separation of pixels $p$ and $p'$ and $w_g(\theta_{pp'}) = \langle \delta_g(p)\,\delta_g(p') \rangle$ is the angular correlation function of the galaxy field averaged over realizations of $\delta_g$ with the systematics held fixed. The last term represents Poisson shot noise, with $\bar{N}_{\rm obs}$ denoting the mean count per-pixel. Moving to harmonic space, where we write the window $W$, the pseudo-$C_\ell$ coupling matrix $M_{\ell \ell'}$, and the templates' pseudo-$C_\ell$ on the mask as $\tilde{C}_\ell^{t_i t_j}$, we have
\begin{equation}\label{eq:powerspec}
  \langle \tilde{C}_\ell^{\rm obs} \rangle
  = (1-\eta_{\rm add})^2 \sum_{\ell'} M_{\ell\ell'}\!\left[W(1+m)\right] C_{\ell'}^{g}
  + \tilde{C}_\ell^{\delta_{\rm sys} \delta_{\rm sys}},
\end{equation}
where we have dropped the shot-noise term for simplicity. The multiplicative field acts as an additional windowing of the field $W \to W(1+m)$, while the first-moment contamination enters as its own term.

For the single-pixel case $p = p'$ and again dropping the shot noise term (see Section~\ref{sec:shotnoise} for details on this term), we have 
\begin{equation}\label{eq:singlepix}
\langle \delta_{\rm obs}(p)^2 \rangle = (1 - \etaadd)^2 w_g(0) [1+m(p)]^2 + \delta_{\rm sys}(p)^2.
\end{equation}
The ensemble mean of $\delta_{\rm obs}$ is $\delta_{\rm sys}$ which sees only the sum of the two contaminant terms, while the variance is modulated by a $(1+m)^2$ term.

\subsubsection{Template parameterization}\label{sec:templatepar}

We expand both mechanisms in the same array of $N_t$ templates $t_i$, normalized
to zero mean over the unmasked pixels, following BMDS24:
\begin{equation}\label{eq:mp}
  m(p) = \sum_{i=1}^{N_t} b_i t_i (p),\quad (1-\eta_{\rm add})m(p) + \delta_A(p) = \sum_{i=1}^{N_t} a_i t_i(p),
\end{equation}
so that $b_i$ is the multiplicative response of template $i$ and $a_i$ its total
first-moment response. A template which acts purely multiplicatively has
$a_i = (1 - \eta_{\rm add})b_i$. If additional additive contaminants are present in the sample, these will tend not to respond to $m$ in the same way as real galaxies, potentially diluting the first-moment response. For a sample with no spurious population, $\eta_{\rm add} = 0$ and $a_i = b_i$; for a purely additive contaminant, $b_i = 0$ with some nonzero $a_i$.
Section~\ref{sec:non-linear} replaces the linear response of Eq.~\ref{eq:mp} with
an expanded function basis, and nothing that follows depends on the response
being linear in $t_i$.

Solving Eq.~\ref{eq:deltaobs} for the true contrast gives the corrected map
\begin{equation}\label{eq:deltag1}
  \hat{\delta}_g (p) = \frac{\delta_{\rm obs} - \sum_i a_i t_i (p)}{(1-\etaadd ) (1 + \sum_i b_i t_i)}.
\end{equation}
Expanding the denominator and dropping terms
of $\mathcal{O}(\etaadd m)$ gives
\begin{equation}\label{eq:deltag}
  \hat{\delta}_g (p) = \frac{\delta_{\rm obs}(p) - \sum_i a_i t_i (p)}{1 + \sum_i b_i t_i(p) - \eta_{\rm add}},
\end{equation}
which is Eq.~4 of WSRS26. The difference, as noted therein, with
the estimator of BMDS24 is the inclusion of the integral-constraint
term $\eta_{\rm add}$.

We note that, in WSRS26's notation
$f_{\rm mult} = (1 - \eta_{\rm add})m$, the denominator of Eq.~\ref{eq:deltag} is
exactly $(1-\eta_{\rm add})(1+m)$, which is the exact inversion of
Eq.~\ref{eq:deltaobs}. However, our $b_i$ are fit in BMDS24's
parameterization which carries no integral constraint prefactor on $\delta_g$,
so $\sum_i b_i t_i$ estimates $m$ rather than $f_{\rm mult}$. Evaluating
Eq.~\ref{eq:deltag} with these coefficients is therefore making an approximation of
the exact denominator to $\mathcal{O}(\eta_{\rm add} m)$, which is second-order
in two quantities which are each themselves small, so we choose to neglect.

Much of this paper operates (and much of the historical literature has operated)
in the multiplicative-only limit $a_i = b_i, \eta_{\rm add}=0$, in which
Eq.~\ref{eq:deltag} reduces to
\begin{equation}\label{eq:deltagmult}
  \hat{\delta}_g(p) = \frac{\delta_{\rm obs}(p) - m(p)}{1 + m(p)}.
\end{equation}
This implies that the correction is not simply a subtraction, but also requires
a reweighting of remaining galaxies and a corresponding
anisotropic modulation of the Poisson variance in the density map pixels. To avoid
numerical pathology due to under-sampled templates in some regions, we impose a
mask $|m| < 0.95$, thereby choosing to mask regions where the remaining galaxy
densities would be modulated by a factor of $\geq 20$. In the regime relevant to
SPHEREx, typical template amplitudes rarely exceed $|m| \sim 0.1$.

Since $1+\delta_{\rm obs} = N/\bar N$, the corrected contrast
$(\delta_{\rm obs}-m)/(1+m)$ may be written as $\frac{N/(1+m)}{\bar N} - 1$.
This implies that we may straightforwardly reweight each galaxy by $w= 1/(1+m)$
to reproduce the corrected map, which are typically taken into account during
the power-spectrum estimation stage downstream. Note that the full joint
additive and multiplicative treatment does not admit this, requiring additional
consideration for downstream analysis.

\subsubsection{Template preprocessing}\label{sec:preprocessing}
Input maps of systematic effects are fed through a series of preprocessing
steps, depending on their format and physical dynamic range. Stellar density and
reddening may span many orders of magnitude and have highly skewed pixel-value
distributions, so these are transformed as $\log_{10}(t+\epsilon)$ for those which are
strictly positive, with a constant $\epsilon$ set below the smallest physically
meaningful value to guard against numerical noise. Templates which are close to symmetric and have more limited
dynamic range (zodiacal brightness, source density/coverage) are standardized to
zero mean and unit variance. The preprocessing configuration is decided on a
template-by-template basis based on inspection and the nature of the physics
involved in constructing said template, rather than something that is decided
on-the-fly by the fitting algorithm. As will be demonstrated in Section~\ref{sec:res-non-lineardet},
the characteristics of our estimator provide some leeway to absorb the impact of
a variety of (potentially less-than-optimal) preprocessing choices here (Section~\ref{sec:res-templatecorr} and Figs.~\ref{fig:binneddensity} and \ref{fig:extinction857}). Templates are also aligned in their coordinate systems (all
aligned to equatorial) and to \texttt{NEST} HEALPix ordering, having been
brought to a common \texttt{NSIDE} resolution beforehand. We never down-sample
the galaxy field, and require any template to be generated natively at, minimum, half of the galaxy
field resolution when possible (some external maps are not provided [or
usefully resolved] at all desired resolutions, so these are up-scaled but used as-is).

\subsection{Template curation and conditioning}\label{sec:curation}
A common problem in template deprojection approaches is the joint handling of
systematic templates which are themselves highly correlated. Many may be
redundant measures of the same effect, or different
systematic mechanisms may be correlated by some underlying physical property, such
as the local overall galactic population/their manifestation in galactic plane,
or instrument observing properties that are jointly affected by elevated transient anomaly rates (often correlated on the celestial sky in SPHEREx's case due to their association with the South Atlantic Anomaly [SAA]). Some templates may contain truly
redundant spatial information and have Pearson correlation coefficients near 1,
while others (many) have lower levels of residual correlated structure.

Part of the aim of this paper is to enable the consideration of a wider ensemble
of templates more easily, but this ``kitchen sink'' approach is at-odds with
wanting to keep template correlations under control. For instance, the nominal
stellar density proxy template tends to be correlated with the dust extinction
template which both trace Galactic structure. Algorithms such as the
ElasticNet are able to consider the full landscape of template-template
covariance in performing their fits, but many realistic test kits exhibit poor
condition numbers, clearly meriting closer
consideration even for algorithms which incorporate covariant fits. These
degeneracies can introduce instability and increase runtimes of estimators
significantly if this is not addressed. We therefore impose a curation step on the raw nominal
template ensemble prior to feeding them to the fitting algorithms. We emphasize
that this is purely a template-template pre-treatment step, which makes no
statements about the correspondence of template features to the data.

The overall notion is to employ Gram-Schmidt (GS) orthogonalization to extract
independent components of our correlated template set. We first start by
hand-selecting a list of templates considered as ``primaries'' -- templates
which are never themselves allowed to be orthogonalized, but whose correlations
with other templates will define those templates' orthogonalizations. This
resolves the slight chicken-and-egg problem of which templates are referenced
against which and ensures that the solutions remain relatively stable under small perturbations in the data or variations in template ensemble, though requires some care in establishing that the primaries
themselves are not significantly correlated. This set of primaries is kept
deliberately minimal, incorporating the templates believed to be the most
relevant and physically motivated, ensuring that their physical interpretations
are preserved. Primaries may still be somewhat correlated with one another, however, motivation the inclusion of a collinearity term in the estimator (Section~\ref{sec:elasticnet}). Three possible actions are then taken on the
remaining templates, depending on the maximum Pearson correlation with respect to the other template ensemble members:

\begin{enumerate}
  \item{\textbf{Aggregate} template families which are duplicates by
        construction. Some simple or intrinsically low-resolution templates will
        genuinely contain no additional unique information, so a non-primary
        correlating with a primary would be dropped, while two secondaries
        correlating would drop the second in the list. We set a nominal
        criterion of $\rho \ge \rho_{\rm dup}$ on the maximum correlation coefficient for a
        given template in order to trigger this mechanism, using a nominal value
        of $\rho_{\rm dup} = 0.99$, though in the
        production context this threshold would be calibrated explicitly based
        on additional simulations.}
  \item{\textbf{Keep whole} templates which are robustly independent of the rest
        of the ensemble, as well as the primaries. We set the nominal threshold
        at a maximum of $\rho \leq \rho_{\rm orth}$, again calibrated to
        simulations but set with an initial fiducial threshold of
        $\rho_{\rm orth} = 0.20$.}
  \item{\textbf{orthogonalize} templates which are moderately-but-not-fully
        collinear. Following and extending a pattern used for dust
        mis-estimation residuals in DESI imaging weights~\citep{chaussidon2022}, a
        template that is correlated with the remaining set above
        $\rho_{\rm orth}$ is replaced
        with its Gram-Schmidt residual against that set. This residual is
        constructed to be zero-mean over the mask and is ingested with no
        additional preprocessing.}
\end{enumerate}

One might reasonably express the concern that removing significant components of
a map would affect the linearity of its overall fit against the data, even if
the orthogonal components were retained, or that template pairs might themselves
have structure in their cross-correlation beyond linear order (Section~\ref{sec:res-templatecorr}). A potential way of addressing this would be to extend the curation and orthogonalization over the full Legendre
basis up to $K=6$, and we demonstrate in Section~\ref{sec:results} that the simple linear treatment would
indeed be insensitive to higher-order template-template effects.
Future analyses may choose to perform the curation procedure over an
$N_{t} \times N_{t} \times N_{K}$ matrix, rather than $N_t \times N_t$, to
account for this.

\subsection{Selection and regression}
\label{sec:selection}

\subsubsection{Separating selection and estimation}\label{sec:separation}

A penalized regression such as ElasticNet (EN, Section~\ref{sec:elasticnet}), faced with the task of template
deprojection, is being asked to perform two distinct operations. The first is
selection, which for EN is done via a ``lasso'' penalty driving the
coefficients of unhelpful templates towards zero. However, as part of this same
mechanism, it will tend to shrink the coefficients of the retained templates,
leading to a failure to recover injected coefficients in mocks and, more
troublingly, leaving residual contamination in the post-deprojected map. This
residual may also have complicated correlating structure with the remainder of
the template ensemble, leading to unpredictable or unintended behaviour.

We may avoid this by separating the step of template \emph{selection} from that
of the actual \emph{estimation}. We still employ the information provided by the
penalized regressors, but instead use their recovered coefficients to define a
binary mask: templates with coefficient amplitudes which are detectable above
the floating-point precision floor are passed directly to an ordinary
least-squares refit to determine their coefficients without shrinkage bias,
while templates with coefficients indistinguishable from zero are culled.

The implications of this are both that coefficients are recovered without
shrinkage bias, and notably that it aligns the estimation procedure behind the two
alternative selection algorithms (EN and IFS). That is, for a curated ensemble
for which the two selectors produce the same binary mask, the actual regression
is performed identically, aiding interpretation of their comparison. This is effectively
the post-lasso estimator of \citet{belloni2013}, which achieves a rate of convergence comparable to
or better than the standard lasso, with smaller bias; this can also be seen as the unpenalized
limit of the relaxed lasso of \citet{meinshausen2007}.

We employ the two selectors, ElasticNet (Section~\ref{sec:elasticnet}) and Iterative Forward Selection (Section~\ref{sec:ifs}), whose consistency informs template basis choices, with two methods of estimation, ordinary least-squares (OLS) and a version of the maximum-likelihood estimator of BMDS24 (Section~\ref{sec:regress}). OLS is a fast and reliable standard, while the MLE estimator offers additional flexibility in accounting for additive systematic terms.

\subsubsection{Stage 1, Selector 1: Information Criterion-scored ElasticNet}\label{sec:elasticnet}

Our primary selector is a variant of the ElasticNet of \citet{zou2005}, which
combines a lasso penalty on the summed absolute coefficients $\|\mathbf{a}\|_1 = \sum_i |a_i|$, which encourages sparsity, with a ``ridge'' penalty on their summed squares $\|\mathbf{a}\|_2^2 = \sum_i a_i^2$, designed to
better account for collinearity. The simple lasso estimator is
unstable under collinearity, and particularly the primary template set may still
retain a moderate degree of correlation after the curation procedure, motivating
us to retain the full ElasticNet rather than the simpler lasso operator. The standard ElasticNet fits coefficients on the set $P$ of the $n = |P|$ unmasked pixels entering the fit by minimising
\begin{equation}\label{eq:enloss}
\begin{aligned}
  \hat{\mathbf{a}}(\alpha, \rho; P) = \argmin_{\mathbf{a}} \Big[ &\frac{1}{2|P|} \sum_{p \in P}
  \Big( \delta_{\rm obs}(p) - \sum_i a_i t_i(p) \Big)^2
  \\
  &+ \alpha\rho\, \|\mathbf{a}\|_1 + \frac{\alpha(1-\rho)}{2}\, \|\mathbf{a}\|_2^2 \Big],
\end{aligned}
\end{equation}
where $\delta_{\rm obs}(p)$ and $\sum_i a_i t_i (p)$ are the respective data vector and total template values in
pixel $p$, $\alpha$ the penalty strength and $\rho$ the fraction carried by the lasso term.

The standard \texttt{ElasticNetCV} as implemented in \texttt{scikit-learn}~\citep{scikit-learn} selects $\alpha$ and $\rho$ by minimising the mean squared
error (MSE) on held-out cross-validation folds (data subsets),
\begin{equation}\label{eq:mse}
  M_k(\alpha, \rho) = \frac{1}{|F_k|} \sum_{p \in F_k}
  \Big( \delta_{\rm obs}(p) - \sum_i \hat{a}_i(\alpha, \rho; P \setminus F_k)\, t_i(p) \Big)^2,
\end{equation}
where $F_k$ and $P\setminus F_k$ are the $k$-th held-out fold and its training complement
($K=5$), respectively, and $\hat{a}_i$ are the elements of $\mathbf{\hat{a}}$. Selecting the penalty by out-of-sample error controls overfitting, but optimizes
for predictive accuracy, when we might instead like to optimize for predictive
parsimony. The former keeps every template that improves its prediction,
including those whose contribution to removing true contamination is negligible with respect to the
cosmological signal that they may remove.

We therefore use the ElasticNet loss only to propose candidate template sets. For each of
$\rho \in \{0.1,\,0.5,\,0.7,\,0.9,\,0.95,\,0.99\}$ and 50 values of $\alpha$ spanning
three orders of magnitude below the value which zeroes every coefficient. Each solution defines one candidate, the set of templates it retains,
\begin{equation}\label{eq:candidates}
  S(\alpha, \rho) = \left\{ i : \hat{a}_i(\alpha, \rho; P) \neq 0 \right\}.
\end{equation}
Each candidate $S$ is refit via unpenalized least squares over all
$n = | P |$ pixels along with an intercept term (Section~\ref{sec:separation}),
scored by
\begin{equation}\label{eq:bic}
  B(S) = n \ln M_{\rm OLS}(S) + \xi |S| \ln n
\end{equation}
with
$M_{\rm OLS}(S) = \frac{1}{n}\sum_{p \in P} \big( \delta_{\rm obs}(p) - \sum_{i \in S} \hat{a}^{\rm OLS}_i t_i(p) \big)^2$,
selecting $\hat{S} = \argmin_S B(S)$. For $\xi = 1$, this is the Bayesian
Information Criterion (BIC) of \citet{schwarz1978} for a Gaussian likelihood,
evaluated at the maximum-likelihood fit of the model $S$. The generalization to $\xi \neq 1$ is the
modified BIC of \citet{wang2009}, and more generally a member of the generalized information criterion 
family~\citep{nishii1984,zhang2010}, with our version differing in that $\xi$ is treated as a calibrated
constant rather than a diverging series. \citet{zou2007} demonstrate that the number of retained nonzero
coefficients is an unbiased estimate of the lasso estimator's degrees of freedom, enabling use of a standard BIC; our
approach differs in that we score the OLS refit of each candidate template set, rather than the shrunk lasso
fit directly. $M_{\rm OLS}(S)$
decreases monotonically as templates are added, which is counteracted by the
complexity-penalising second term, whose aggressiveness is set by the penalty
scale $\xi$. In spite of this generalisation, we may refer to this estimator as
a BIC occasionally herein for convenience.

The tunability of the scale $\xi$ is a distinct advantage of this approach.
The na\"{\i}ve BIC penalty $|S| \ln n$ for a number of observations $n$ assumes that all $n$ observations are independent; for a realistic and coherent
effect on the sky, our $\geq \mathcal{O}(10^5)$ pixels are spatially correlated on
scales much larger than the individual pixel sizes, so the number of true
independent modes is orders of magnitude smaller, and the default $\xi = 1$
scale factor is nearly ineffectual. Rather than imposing assumptions about the
correlation scale, which would presuppose the nature of the structure we
are trying to characterize, we instead treat it as a calibrated parameter, maximizing the recovery of a weak injected signal while
exhibiting a false-positive rate $< 5\%$ as assessed by contaminated and
uncontaminated mocks, respectively. We set this as a fiducial
criterion for early analyses which we believe to be reasonable given the size of the overall
ensemble, but for the ultimate $\fnl$ results we plan
to calibrate these thresholds and our overall null test and unblinding criteria
with respect to their downstream impact. For the calibration of a production analysis,
this requires a careful matching of the
abundances and sampling of the mocks and the input data catalogue. We may also interpret it directly as a
per-template significance threshold. Adding one column (template) of $t$-statistic $t$ changes the score by
$\Delta B \simeq -t^2 + \xi \ln n$, so a template enters when
$ |t| \gtrsim \sqrt{\xi \ln n}$. For the $n=560{,}496$ fitted un-masked pixels of the
mocks described in Section~\ref{sec:mocks}, this corresponds to $|t|$ of 3.6 at
$\xi = 1$, 12.6 at $\xi = 12$, and 17.8 at $\xi = 24$.

\subsubsection{Stage 1, Selector 2: iterative forward selection}\label{sec:ifs}

The second selector, used as a cross-reference for the primary ElasticNet, is a variant of the Iterative Forward Selection 
approach detailed by WH21. IFS is a greedy algorithm that regresses one template
at a time, based on a metric of which template of the remaining set has the most
information in common with the data. This template is then regressed
individually and removed from the list, after which the procedure is iterated
until the list is empty or the stopping criterion is reached. The selection path
is ordered, providing useful diagnostic information by reflecting which
templates were most important, but may struggle with template pairs which are
highly correlated (though the two-step OLS procedure softens this significantly
for the estimation step).

Our baseline implementation differs from the specific one described in WH21 in that we
again utilize a $\Delta B$ criterion akin to Eq.~\ref{eq:bic}
to determine which template
is accepted at a given step, rather than the covariance of the fit's own
parameters, and similarly use a calibratable BIC-based stopping criterion,
thereby making the two selectors conceptually comparable under a common choice
of $\xi$. We note, however, that a common choice of $\xi$ does not
necessarily result in a common calibration/false-positive rate. We also choose
to remove templates from the list upon selecting them, rather than choosing with
replacement. We have also implemented and present the WH21 implementation which
instead bases its fits and stopping criterion on the analytic covariance of the
fitted parameters, though we show this only for comparative purposes. 

We also note a third, non-adaptive specification used solely as a reference on
mocks. The matched control operator is defined with respect to contaminated
mocks as the fixed list of templates used to contaminate, with no data-driven
selection. Because this set applies the same operation to every map, it isolates
the transfer function from the selection, and directly tests the
intercompatibility of the contamination and deprojection models (e.g., the
effect of applying a multiplicative deprojection model to a map containing
additive contaminants).

\subsubsection{Stage two: least squares, or the joint
  likelihood}\label{sec:regress}
The default stage is an OLS regression as described above,
which performs a joint regression of the templates selected by Stage~1, and
estimates one amplitude per template (column of the ``template matrix''). By
Eq.~\ref{eq:deltaobs} this first-moment estimator sees only the combination
$a_i$, and the correction that it constructs via Eq.~\ref{eq:deltagmult} is
multiplicative by construction and so can only correct the mean shift of $a_i$.
Some physical effects, most notably stars, will contribute both additively and
multiplicatively with an extremely similar spatial pattern, so a simple
regression would fit the mean but misattribute the effect on the variance. In general, one would like to estimate $a_i$
and $b_i$ jointly, which requires a different estimator.

In the single-template case of Eq.~\ref{eq:mp}, $a$ shifts the mean of
$\delta_{\rm obs}$, and is therefore recoverable by ordinary regression, while
$b$ shifts the variance, and so is invisible to estimates of the first moment (Eq.~\ref{eq:singlepix}).
To handle this, BMDS24 solve for them by writing a joint
likelihood which we adopt, leveraging the fact that the different components are constrained by
different moments of the map. Under a Gaussian model for $\delta_g$ with pixel
variance $\sigma_c^2$, and writing the residual
$r = (\delta_{\rm obs} - at)/(1+bt)$, the negative log-likelihood is
\begin{equation}\label{eq:berlfeinlikelihood}
  - \ln \mathcal{L} = \frac{N}{2}\ln 2\pi \sigma_c^2 + \sum_p \ln | 1 + b t(p) | + \frac{1}{2\sigma_c^2} \sum_p {r(p)}^2,
\end{equation}
where the information on $b$ is carried by the Jacobian middle term arising from
the change of variables from $\delta_g$ to $\delta_{\rm obs}$. Treating pixels as independent ignores the off-diagonal clustering terms of Eq.~\ref{eq:2pt}, but as the likelihood only depends on each pixel's mean and variance (Eq.~\ref{eq:singlepix}), the estimates remain unbiased. This comes at the price of larger scatter and unreliable uncertainties, which we do not use, instead estimating the fluctuation from the mock ensemble.

In the alternate mode, the Stage 2 least-squares refit is replaced outright by
the likelihood in Eq.~\ref{eq:berlfeinlikelihood}, though the Stage~1 selection
and basis expansion machinery is still leveraged as-is. This aggressive
selection is crucial for reducing the number of degrees of freedom in the
ultimate correction. We also
note that the term ``deprojection'' has stopped applying to this procedure in
the strict sense, but we retain use of the term for its familiarity and
applicability to the OLS option.

One potential subtlety at this stage is the fact that the Stage~1 selection via
the BIC-informed ElasticNet or IFS is only sensitive to the first moment of the
map, and one might be concerned that it would not detect maps whose signal is in
the second. As noted in Section~\ref{sec:templatepar}, physical effects will
generally not have $a_i = 0$ with $b_i \neq 0$; the exception is the (somewhat
fine-tuned) example of exact cancellation between the two mechanisms for a given
effect.

\subsubsection{Shot noise contribution}\label{sec:shotnoise}

The pixel variance in Eq.~\ref{eq:berlfeinlikelihood} is attributed entirely to
the modulated clustering term, so the variance of $\delta_{\rm obs}$ scales as
${(1+b t)}^2$. However, shot noise does not scale this way, so an unmodulated
variance is read by the likelihood as a weaker overall fluctuation and $b$ will
tend to be attenuated by an amount set by the shot-noise fraction of the total
pixel variance.

At BMDS24's number densities, this effect
is negligible (shot fraction $\sim 0.1$), but cannot be ignored for SPHEREx's
number densities and pixel counts where the shot contribution is significant.

Writing the contrast $N(p)/\bar{N} - 1$, its Poisson variance is
$\mathbb{E}\left[N(p)\right]/\bar{N}^2$, and marginalising over the zero-mean
true field leaves $\mathbb{E}\left[ N(p) \right] = \bar{N} (1 + at)$. The shot noise
term then follows the mean density, and so is set by the additive amplitude,
whereas the clustering term is modulated by the multiplicative one:
\begin{equation}\label{eq:shotnoise}
  v(p) = \sigma_c^2 {[1 + b t(p) ]}^2 + \frac{1 + a t(p)}{\bar{N}}.
\end{equation}
The two terms scale differently in the case that $a \neq b$, and we see that the
high-density limit $\bar{N}^{-1} \to 0$ recovers the middle term from
Eq.~\ref{eq:berlfeinlikelihood} identically. The skew-normal form presented by
BMDS24 admits no closed-form convolution with a Poisson term, but we
find it unnecessary in any event as the residual of our mock field is close to
symmetric.

We note that a limitation of the joint fit is that the mean additive
fraction $\eta_{\rm add}$ by its nature cannot be fit from the pattern of the
map itself, and must be reconstructed or estimated based on external
information. WSRS26 for instance pursue this by attempting to
estimate the misclassification rate of the star/galaxy separation procedure.

\subsection{non-linear basis expansion}
\label{sec:non-linear}

\subsubsection{Construction}\label{sec:nlconstruction}

The preceding sections have assumed that the induced variations in
density respond linearly to the values in the spatial templates. One may immediately
intuit some regimes where this assumption may be dubious: stellar density and
dust extinction in the most extreme regions; the aggregate of the contaminating
and obscuring components of stars if doing a purely-multiplicative treatment;
and density discontinuities at forced photometry reference catalogue survey
boundaries will all couple to the density field in ways that may have more
complexity and variation than a simple linear scaling would allow. This is also,
undesirably, a function of the preprocessing described in Section~\ref{sec:preprocessing}. We
would like to free ourselves from these simple assumptions, and mitigate the risk
of suboptimal choices at the preprocessing step, by allowing for additional
degrees of freedom in the fit.

To do this, we expand each template into a basis prior to the
selection/regression steps, replacing Eq.~\ref{eq:mp} with
\begin{equation}\label{eq:basis}
  m(p) = \sum_{i=1}^{N_t} \sum_{k=1}^{K} b_{ik} \phi_k (x_i (p)),
\end{equation}
where $x_i$ is the preprocessed template rescaled to $[-1, 1]$ and $\phi_k$ is
the $k$th basis function (note that this is distinct from the $k$ used to label
cross-validation folds in Section~\ref{sec:elasticnet}). By expanding the
template in this way, we allow for the model to be non-linear in the template
values while remaining linear in the coefficients, allowing us to leverage the
existing body of linear-regression machinery including the selection and
two-stage regression machinery from Section~\ref{sec:selection} unchanged. This also has the advantage of making the coefficients directly interpretable, with $b_{i1}$ the linear
component of the response for template $i$, $b_{i2}$ the leading curvature, and so on.

The default basis we employ is one of Legendre polynomials $P_k$,
$k = 1\dots K$, which are orthogonal on $[-1,1]$ such that, in the idealized
case of a uniformly distributed template\footnote{The fact that the empirical
  distribution of template values is not uniform will cause degradations of the mode orthogonality in practice.}, each order is fit independently such
that the selection on order $k$ is not affected by that on order
$k'$. The
zeroth order is omitted, being degenerate with the fitted intercept. The rescale to $[-1,1]$ is done by
percentile clipping at 1\% and 99\% to avoid small numbers of pixels compressing
the dynamic range of the basis. Legendre polynomials are the default, but an alternative basis option
B-splines of specified degree and knot count. This high degree of
configurability of the B-splines leads us not to attempt to optimize over an ensemble of many
orthogonal components, but rather to deliberately place spline knots at the
locations of known sharp/localized features. The B-spline option complements the default Legendre basis, being appropriate to apply selectively in specialized situations.

\subsubsection{Automatic order determination}\label{sec:orderdet}

The non-linear expansion, automatic selection, and orthogonality of our template
treatment so far naturally compose to create a framework where
template-order components are determined directly by the data. The columns of the template matrix being solved are now the individual basis components of all
templates, and so each individual component must now justify its inclusion
against the data based on the BIC complexity penalty of Eq.~\ref{eq:bic}. A
template whose true coupling to the observed overdensities was purely linear
would retain only its $P_1$ component, and a template whose coupling was
quadratic would retain only its $P_1$ and $P_2$ components, with higher order
terms excluded by the penalty. The maximum Legendre order $K$ is therefore an upper bound
of the deprojection order, with the actual order determined by the data.

This procedure leverages non-linearity where it is supported by the data, while
avoiding a multiplicative explosion of deprojected components. For instance,
$K=6$ with a set of 7 templates will present the selector with 42 components
from which to choose, but if only 1 of the 7 has and is selected as having
non-linearity, then only a few extra free parameters are necessary, rather than
the full 42,
and the cosmological signal of interest does not pay the penalty for the degrees of
freedom that the data did not choose (see Section~\ref{sec:res-transfer}).
This is what fundamentally distinguishes this
approach from the neural network or gradient-boosted tree non-linear deprojection
methods employed elsewhere in the literature. The retained order is itself also a
directly interpretable measurement about which systematic effects have non-linear
couplings to the observed overdensities, and at what order, potentially
informing future analyses' approaches to preprocessing or fitting even if this
full framework is not employed. We demonstrate these features of the framework
in Section~\ref{sec:res-non-lineardet}.

\subsubsection{Composition with the joint fit}\label{sec:jointcomp}

The basis expansion naturally composes with the joint Stage-2 estimator of
Section~\ref{sec:regress} without modification. The per-template model of
Eq.~\ref{eq:mp} remains linear in $\{a,b\}$ under any transform of the template,
a point that BMDS24 themselves make explicitly in motivating their own
restriction to linear terms in $t$. Our Legendre columns simply act as
additional columns in the fit, each carrying its own additive or multiplicative
amplitude, and so the composition recovers a joint non-linear multiplicative
response.

\section{Mocks and validation framework}
\label{sec:mocks}

\subsection{Mock catalogues}
\label{sec:mock-cats}

For validation, we use 100 lognormal galaxy catalogues generated using the public
\textsc{Julia}~\citep{bezanson2017julia} package \texttt{LogNormalGalaxies.jl}\footnote{\url{https://github.com/hsgg/LogNormalGalaxies.jl}}.
Each realization is built from seven independent lognormal boxes, one per
redshift bin, partitioning the range $0 < z < 2.2$ according to the bins
presented in the SPHEREx baseline forecasts~\citep{heinrich2024}. The number density in each bin is
set by the forecasts for sources measured at high enough signal-to-noise to
enter the SPHEREx redshift sample ($\sigma(z)/(1+z) < 0.2$), and the linear bias
from the density-weighted forecast bias (1.03 at $z_{\rm eff} = 0.1$ to 3.29 at
$z_{\rm eff} = 1.9$). The input matter power spectrum is generated by CAMB~\citep{camb} at the bin's effective redshift using a \textit{Planck} 2018
cosmology~\citep{planck2018cosmo} with $\fnl=0$.

The catalogues are generated with grid spacings of 8 to 16 $h^{-1}$Mpc, resulting
in Nyquist wavenumbers of 0.20 to 0.40 $h$Mpc$^{-1}$ across bins. Catalogues are
generated on a \textit{Planck}-derived Galactic plane mask covering
$f_{\rm sky} \sim 71\%$ at resolution \texttt{NSIDE} 128. For analysis, we work
on this same footprint at a resolution of \texttt{NSIDE} 256 (yielding
$560{,}496$ unmasked pixels), with each
realization projected to a single angular map.

\subsection{Template kits}
\label{sec:template-kits}

\begin{table}
  \centering
  \caption{\small Template kit. Preprocessing is applied before mean-subtraction over
    the mask.
  ``GS residual'' is the rms amplitude of the template remaining after removing
  the best-fitting linear combination of the templates preceding it, relative to
  its original amplitude. ``Primaries'' are always retained in their entirety.}
  \begin{tabular}{p{0.46\columnwidth}lc}
    \toprule
    Template & Preprocessing & GS residual \\
    \midrule
    Extinction, CSFD $\EBV$ \citep{csfd} & $\log_{10}(t + 0.01)$ & primary \\
    Stellar density, \textit{Gaia} DR3 $12<G<17$ \citep{gaiadr3} & $\log_{10}(t + 1)$ & primary \\
    Ecliptic-latitude depth proxy & standardized & primary \\
    HI4PI $N_{\rm HI}$ \citep{hi4pi} & standardized & 0.39 \\
    \textit{Planck} 857\,GHz \citep{planck2018hfimaps} & standardized & 0.82 \\
    \textsc{Commander} synchrotron \citep{planck2018compsep} & standardized & 0.87 \\
    \textsc{Commander} free-free \citep{planck2018compsep} & standardized & 0.80 \\
    \bottomrule
  \end{tabular}
  \label{tab:kit}
\end{table}

\begin{table*}
  \centering
  \caption{\small Configuration of contamination applied to the contaminated mocks.
    Each configuration is painted at $5\,\sigclean$ and at $2\,\sigclean$ onto the
    same 100 clean realizations. Multiplicative dust extinction and stellar
    density are applied with negative response coefficients (more systematic
    $\to$ fewer sources), while the depth proxy is painted with a positive
    coefficient (deeper map $\to$ more sources). Additive stellar contamination,
    happening at the catalogue level, is a positive contribution of sources, by
    construction.}
  \begin{tabular}{lll}
    \toprule
    Config & Response & Used in \\
    \midrule
    Multiplicative & linear in each preprocessed template & Figs.~\ref{fig:methodcomp}--\ref{fig:tradeoff-contamfp}, \ref{fig:extinction857} \\
    non-linear & extinction as $P_1 + P_3$ of the rescaled template, equal
                amplitudes & Figs.~\ref{fig:ordersel}--\ref{fig:recovery}, \ref{fig:transferfunc} \\
    Step edge & depth proxy as a step at the median of the rescaled template & Fig.~\ref{fig:stepedge} \\
    Additive + multiplicative & multiplicative configuration plus additive stars, 1 per cent of the sample, rate $\propto$ raw stellar density & Sec.~\ref{sec:res-addmult} \\
    \bottomrule
  \end{tabular}
  \label{tab:contam-configs}
\end{table*}

We make no claims about the precise selection or nature of templates which will
eventually be used in SPHEREx's cosmology analysis, but for this mock-based
analysis employ a kit of systematic templates based on external measurements,
plus an analytic ecliptic-latitude depth proxy, shown in Table~\ref{tab:kit}.
The depth proxy is calculated as $t = 1/\max (\cos \beta, \cos 80^\circ)$,
intended to roughly sketch SPHEREx's observing depth as defined by the
combination of its ecliptic scans and the zodiacal emission which sets the local
photon noise level (initially, an external zodiacal emission template was
included as well, but this was highly degenerate with this analytic proxy and most of the orthogonalized residual was contamination from bright stars and galaxies, so this template was excluded from subsequent analysis).

We employ two versions of this kit in the sections that follow. The
``raw'' version passes each template to the selection step as-is, while the
Gram-Schmidt kit applies the orthogonalization of
Section~\ref{sec:curation}, having designated the dust extinction, stellar
contamination, and depth-proxy templates as ``primaries'' that enter unchanged
and against which the other templates (HI, 857 GHz dust emission, synchrotron,
and free-free emission) are orthogonalized. The condition number for this system
is 1.70, with the largest remaining correlation being 0.24 (between extinction
and stellar density). The comparison between these two kits is then reflective of the
change in performance of the cleaning algorithms when larger template-template
correlations are present. This kit contains templates that measure slightly different aspects of the same underlying phenomenon, with the HI map, CSFD extinction map, and 857 GHz emission-tracing map all tracing the Galactic interstellar medium; these will tend to correlate highly, but this orthogonalization and selection process will allow the useful features of all of the maps to be distilled.
  
\subsection{Contamination model}
\label{sec:contam}

Contamination is painted onto the clean catalogues at the per-galaxy level,
ensuring self-consistent shot noise in the contaminated
map after pixelisation. The multiplicative component modifies the probability of retaining a galaxy
by $\max(0, 1+m(p))$ as evaluated in that pixel $p$, with $m(p)$ constructed per
Eq.~\ref{eq:basis}. The optional additive component draws additional catalogue
objects from a Poisson distribution with a rate proportional to the additive
template value, placed at pixel centres and uniformly in volume along the radial dimension. A unique identifier for each configuration together with the realization index defines two deterministic random seeds, one for the per-galaxy retention draws, and one for the additive Poisson draws, ensuring that different configurations painted onto the same realization use different random draws.

Our default multiplicative configuration injects multiplicative dust reddening, stellar
obscuration, and depth proxy terms, with the additive variant containing an additional additive stellar contaminant
term. Injected amplitudes are defined in units
of $\sigclean$, the scatter of each template's OLS coefficients over the 100
clean realizations when all seven raw-kit templates are fit, short-circuiting
the selection stage. These amplitudes are $8.12 \times 10^{-3}$ for dust
extinction, $1.46 \times 10^{-3}$ for multiplicative stellar contamination, and
$9.67 \times 10^{-4}$ for the depth proxy. This floor is set by chance alignment
of the mock realization with the systematic template, which for the large scales
of interest for $\fnl$ are well above the shot noise. Extinction and
stellar density are injected with negative coefficients, as these tend to remove
objects; depth is positively correlated with recovered density and so is
injected with a positive coefficient, so the recovery of the different signs is also
under test. For context, a rough estimate based on their respective low-$\ell$
power suggests that a $1\,\sigclean$ injection of the dust extinction template
used here would account for the full $\delta n / \bar{n}$ budget for SPHEREx's
target $\fnl$ sensitivity, with the depth and stellar density templates accounting for about half and one sixth of the budget per $\sigclean$, respectively. This depends on the chosen template patterns, but serves to give a notion of the scale and requirements for cleaning. We inject each with amplitudes of
$5\,\sigclean$ and $2\,\sigclean$, chosen to probe multiple signal-to-noise regimes. At
$5\,\sigclean$, the multiplicative field spans $-0.107 \leq m \leq 0.048$.

A smooth contaminant with a hard step, as might be the case for effects like
discontinuous imaging depth, is applied at the corresponding amplitude at
ecliptic latitude $|\beta|=27.9^\circ$ (the 50th percentile of the depth proxy). A template with a step could of course be used instead and a linear relationship maintained; this setup supposes some known non-linearity in how the underlying density field responds to one's template map, which we here take to be along ecliptic latitude as a demonstration. Additive stellar
contamination is incorporated proportional to raw star counts (injected so as to
mimic a $\sim$1\% global mis-classification $\etaadd = 0.00995$), while the kit
from Section~\ref{sec:template-kits} is configured to deproject in units of
$\log_{10}$ star counts, deliberately baking-in a non-linear relationship between
the response and the fitted coordinate. The contaminating
configurations, and the figures they apply to, are shown in
Table~\ref{tab:contam-configs}.

\subsection{Figures of merit}
\label{sec:fom}

\begin{enumerate}
  \item{\textbf{Selection.}
        A template is counted as selected when any of its columns (Legendre
        orders) is retained by the selection step. The clean false positive rate
        is the frequency of selection on uncontaminated mock maps, averaged over
        the four never-injected templates. For contaminated configurations, we report the selection
        rates of both the injected and uninjected (false-positive) templates.
        }
  \item{\textbf{Coefficient recovery.}
        We also report the extent to which the injected contaminant coefficients
        are recovered numerically. For linear, multiplicative contaminants, this
        amounts to the correct recovery of the injected slope. Recovery may be
        expected to differ in the raw and GS cases, where in the latter part of
        an injected template may be projected onto another component. We report
        unconditional means, counting unselected coefficients as zero to avoid
        bias due to the ``winner's curse''.
        }
  \item{\textbf{Map-level Residuals.}
        The map-level residual statistic is the mean density contrast over ten equal-pixel-count
        bins in the template value (the binned statistic as in WH21 Figure~1).
        We score each of these with a Mahalanobis
        $\chi^2$~\citep{mahalanobis1936} against similarly processed uncontaminated mock
        realizations, utilising a Ledoit-Wolf~\citep{ledoitwolf2004} estimator, which shrinks the sample covariance towards a scaled identity, with an intensity estimated from the realizations, to mitigate the too-few-realization penalty. This penalty is
        particularly severe for the full $K=6$ non-linear case, so we report each
        method's $\chi^2$ in units of its clean-distribution median, and set no
        threshold on the absolute $\chi^2$. Scoring is done
        on a leave-one-out basis, removing a contaminated realization's clean
        parent realization from the null ensemble.
        }
  \item{\textbf{Transfer Functions and residual power.}
        The concern that gave rise to the information criterion-calibrated
        penalties was the loss of cosmological signal due to overly aggressive, inadvertent ``mode removal'', which
        we assess by comparing the pre- and post-deprojected clean mock pseudo
        power spectra, $C_\ell$. We deproject each clean realization with the
        set of templates selected on its contaminated counterpart. The resulting signal transfer function for the deprojection operation
        $T(\ell) = \langle C_{\ell}^{\rm clean,corr} \rangle / \langle C_{\ell}^{\rm clean}\rangle$
        is then an estimate of the cosmological power removed by the cleaning.
        We estimate the residual power as the injected multiplicative field minus
        the difference between the contaminated and clean model fits to the same
        mock realization, which includes injected structure that the basis cannot
        represent. This includes the effect of Poisson noise on the fitted coefficients, but not the Poisson noise of the maps themselves.
        For the \texttt{NSIDE}=256 maps, the maximum usable
        $\ell$ is $3\times\texttt{NSIDE}-1=767$. We work in 8 logarithmically-spaced bins in $\ell$ and
        use only a simple correction for $f_{\rm sky}$, with no apodisation or
        accounting for mode coupling. As the quantities shown are power spectrum
        ratios, they are relatively agnostic to the details of these choices.}
  \item{\textbf{Second-moment attribution statistic.}
        The map-level statistics above are first-moment in the templates, and so
        are not able to distinguish additive and multiplicative contaminants
        with the same spatial pattern. Treating an additive contaminant as
        multiplicative results in a residual term
        $\hat\delta_g - \delta_g \simeq -a\,t\,\delta_g$, a modulation of the
        clustering amplitude by the template. We use the clean realization
        paired with each contaminated map and regress
        $\hat\delta_g - \delta_g$ on $t\,\delta_g$ over the fit mask, where $t$ is the
        additive contaminant's spatial template, and normalize the slope by the
        fitted $-a$ so that 1 corresponds to the term being fully misattributed
        as multiplicative, and 0 corresponds to it being correctly attributed as
        additive. Because this requires knowledge of the true clean field, it
        applies only to mocks.
        }
\end{enumerate}

\section{Results}
\label{sec:results}

\subsection{Selection comparison and false-positive control}
\label{sec:res-selection}

\begin{figure}
  \includegraphics[width=\columnwidth]{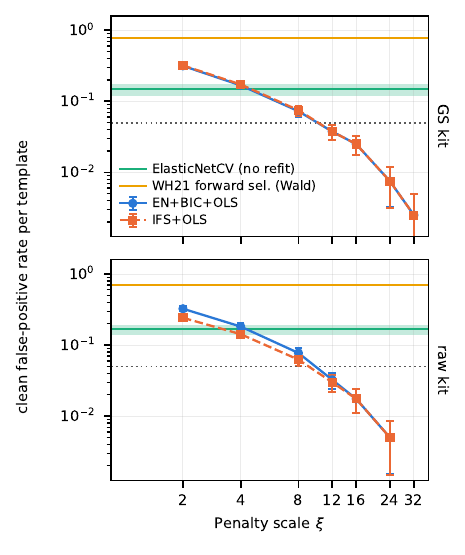}
  \caption{\small Comparison of published methods' false-positive rates, as
    assessed from uncontaminated mocks, against the two hybrid methods presented
    in this work, as a function of their selection penalty scale $\xi$. The
    standard \texttt{ElasticNetCV} is shown in green with its mean and
    $\pm1\sigma$ spread over realizations, and the WH21 forward-selection is
    shown in yellow-orange (its small spread relative to its mean not visible at
    this scale), while the EN+BIC+OLS and IFS+OLS hybrid methods are shown
    respectively in blue circles and red-orange squares as functions of $\xi$
    along with their $\pm 1\sigma$ errors. The fiducial 5\% criterion is shown
    as a dotted black horizontal line, achieved at scale $\xi=12$. Unresolved
    cases (no false positives) are not shown.}
  \alttext{Two plots, one atop another, showing clean-mock false positive rate as a function of the penalty scale $\xi$, and showing flat horizontal lines for the non-tuneable literature \texttt{ElasticNetCV} (green) and WH21 IFS (yellow) methods. Both the EN+BIC+OLS (blue) and IFS+OLS (orange) methods presented here cross-over the \texttt{ElasticNetCV} line beyond $\xi$ of around 4, with error bars extending to a few percent in either direction by $\xi$ of 16 and continuing to increase in size on the log scale as the false-positive rate drops, down to false-positive rates below a percent by $\xi=24$. The WH21 IFS estimator sits well above all others, with a false-positive rate 0.7--0.8. The top panel shows the Gram-Schmidt orthogonalized template set, while the bottom shows the non-orthogonalized set, where the EN+BIC+OLS and IFS+OLS selectors begin to diverge by a few percent in false-positive rate at low $\xi$, IFS+OLS being the better-performing of the two; the GS and raw kits perform similarly overall by this metric, with the raw kit being slightly lower.}
  \label{fig:methodcomp}
\end{figure}

\begin{figure*}
  \includegraphics[width=\textwidth]{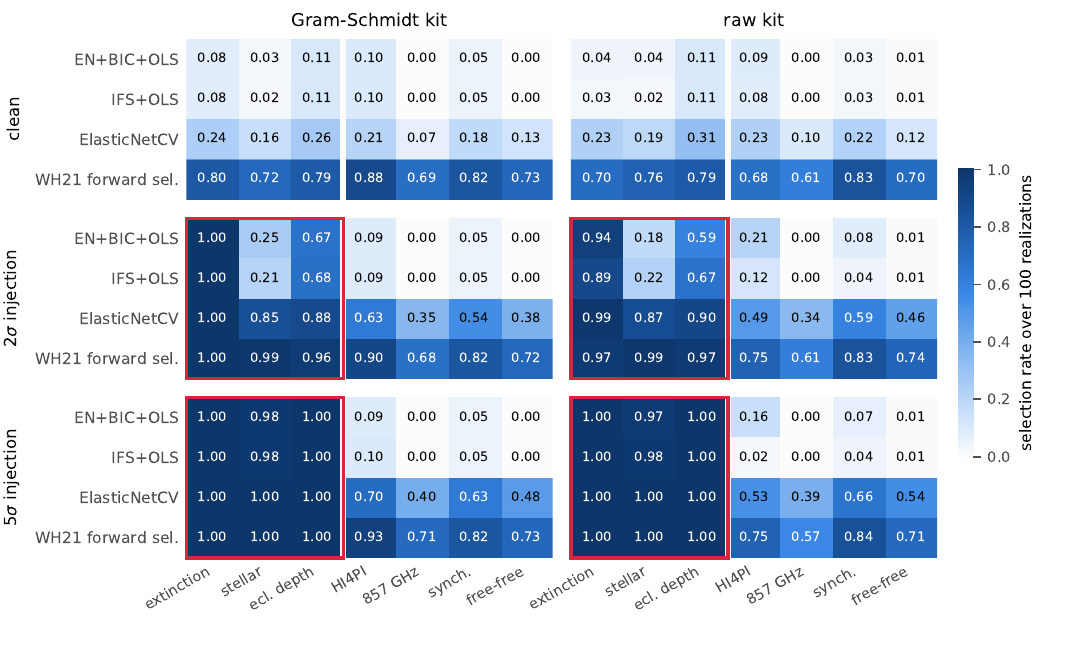}
  \caption{\small Selection rates (as the fraction of 100 mock realizations) for
    clean, low-contamination-amplitude, and
    high-contamination-amplitude mocks, for the Gram-Schmidt orthogonalized
    and raw template kits. Within each block, the first three columns
    (extinction, stellar density, and ecliptic depth proxy), are injected in the
    2 and $5\,\sigclean$ cases, highlighted in red, while the remaining four columns represent templates which are provided to the regression but were never injected (false positives). The EN+BIC+OLS and IFS+OLS hybrid methods are generally much less susceptible to false-positives, particularly in the presence of other contamination, but trade a loss of recovery of some weakly-injected patterns.}
    \alttext{A matrix of selection rates for various estimators, individual templates, template kits, and amplitudes. The figure is arranged as a 3-row by 2-column array of sub-matrices, with the three rows being, from top to bottom, the clean, low-amplitude, and high-amplitude contaminant injections, while the left column is the Gram-Schmidt orthogonalized template kit and the right is the raw kit. Within each sub-matrix, the four selectors, EN+BIC+OLS, IFS+OLS, \texttt{ElasticNetCV}, and WH21 IFS, span the vertical axis and the seven templates span the horizontal: dust extinction, stellar density, and the ecliptic depth proxy (all injected), and the HI4PI, \textit{Planck} 857 GHz, \textit{Planck} \textsc{Commander} synchrotron, and \textit{Planck} \textsc{Commander} free-free emission maps (provided to the selectors, but not injected). The WH21 has high selection rates for injected and uninjected templates throughout. In all cases, the calibrated EN+BIC+OLS and IFS+OLS selectors have visibly lower false-positive rates than the other estimators, by factors of a few and by a large factor when additional contamination is present. For the strong injections, all selectors recover essentially all injections. For the weak injections, the calibrated selectors struggle to select the stellar density contaminant.}
  \label{fig:selectionrates}
\end{figure*}

\begin{figure*}
  \includegraphics[width=\textwidth]{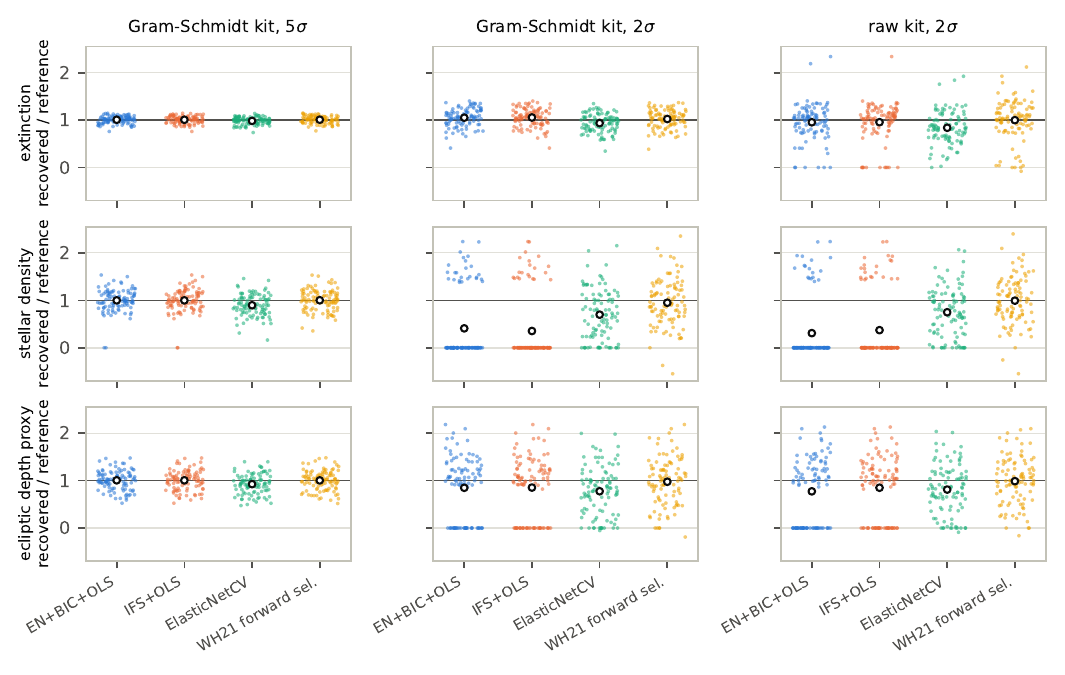}
  \caption{\small Recovery of injected multiplicative coefficients $b_i$.
    From left to right, columns show the $5\,\sigclean$ GS kit, $2\,\sigclean$ GS kit, and
    $2\,\sigclean$ raw kits; from top to bottom, rows show the ratio of recovered
    to injected coefficient for dust extinction, stellar density, and the
    ecliptic depth proxy. The coefficients for \texttt{ElasticNetCV} are biased low in all cases as a result of that estimator's coefficient shrinkage, while both hybrid estimators and the WH21 forward selection operator are unbiased in their injection amplitude recovery for the high signal-to-noise $5\,\sigclean$ injections. For the low signal-to-noise injections, the hybrid selectors set coefficients to zero more aggressively, biasing the mean low (particularly for the stellar density case).}
    \alttext{A 3$\times$3 matrix of estimator-comparing coefficient-recovery scatter plots. Each plot shows the ratio of the estimated to the input contaminant coefficient for each mock realization, with separate scatter plots for each selector. The columns show, from the left, the Gram-Schmidt kit at $5\sigclean$ injected, the GS kit at $2\sigclean$, and the raw $2\sigclean$ case. The rows show, from the top, the recovery of the coefficients for dust extinction, stellar density, and the ecliptic depth proxy. For the GS $5\sigclean$ case, the coefficient is recovered well in most cases by all estimators, with dust extinction being particularly tightly clustered around a ratio of 1. The \texttt{ElasticNetCV} points are biased slightly low in the mean. For the weak case, the coefficients are much more spread-out, the \texttt{ElasticNetCV} downward bias is more apparent, and many of the coefficients in the low-side tail of the calibrated two-stage estimators have been explicitly zeroed, truncating those distributions of points at some minimum value. The distributions are similar in the raw $2\sigclean$ case as in the GS, with the extinction template in particular having visibly more scatter in all estimators.}
  \label{fig:coeff-recovery}
\end{figure*}

\begin{figure*}
  \includegraphics[width=\textwidth]{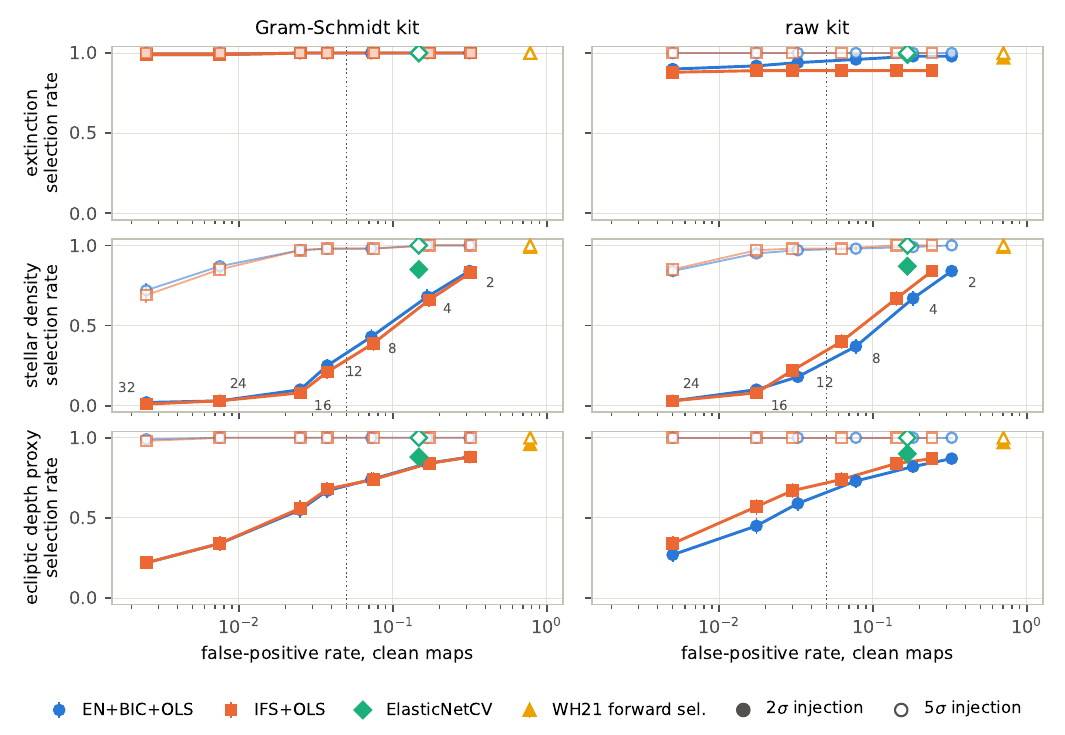}
  \caption{\small Selection rate plotted as a function of clean-mock false-positive
    rate, for the strong (faded, open-symbol) and weak (solid, closed-symbol)
    injections, for both kits and all three linearly injected templates. The
    EN+BIC+OLS and IFS+OLS hybrid estimators are again shown as functions of
    $\xi$ in blue and
    orange-red, respectively (with the $\xi$ values annotated in the middle row), as are \texttt{ElasticNetCV} in green and WH21 IFS
  in yellow-orange. $\pm1\sigma$ error bars over the 100 realizations are
  plotted but are generally small with respect to the values.
  The dotted black vertical line represents the 5\% fiducial threshold.}
  \alttext{A 3-row by 2-column matrix of plots of injected-template selection rate versus clean-mock false positive rate, as a function of penalty $\xi$. The rows show, from the top, the extinction, stellar density, and ecliptic depth proxy contaminants, while the left column shows the Gram-Schmidt orthogonalized template kit and the right the raw kit. Extinction is detected robustly by all estimators at all penalty scales in the GS kit, with the calibrated selectors EN+BIC+OLS (blue) and IFS+OLS (orange) being slightly below 100\% recovery for the raw kit. The strong injections are recovered well at all $\xi$ for the depth proxy for both calibrated estimators, but fall to about 0.7 at higher $\xi$ for the stellar density contaminant. For both the weak stellar density and weak ecliptic depth proxy cases for both kits, 
  the trend is a general slope upwards of increasing false-positive rate with increasing selection efficiency, passing slightly below the \texttt{ElasticNetCV} (green) in selection at a matched false-positive rate, while WH21 IFS (yellow) is far to the top-right with a higher selection rate and false-positive rate than all.}
  \label{fig:tradeoff}
\end{figure*}
Except where noted, the results in this section refer to the linear ($K=1$ in the terms of Eq.~\ref{eq:basis})
design, with the non-linear $K=6$ selector treated separately in
Sections~\ref{sec:res-non-lineardet} and \ref{sec:res-transfer}.

The results of the $\xi$ information-criterion (Eq.~\ref{eq:bic}) sensitivity calibration for the
linear ($K=1$) selector are shown
in Figure~\ref{fig:methodcomp}, and we report results for both the Gram-Schmidt and raw (unorthogonalized) template kits. The false-positive rate as assessed from uncontaminated
mocks falls smoothly with $\xi$ as expected, from 0.32 per template at $\xi = 2$
to $0.003$ at $\xi = 32$ on the GS kit. The fiducial 5\% threshold is met at
$\xi = 12$ (3.75\% GS, 3.25\% raw), so we adopt this value for the linear design. For
the Legendre $K=6$ basis and the five-knot B-spline basis, this is achieved at
$\xi = 24$. For the linear basis at $\xi = 12$, the two hybrid selectors choose
identical sets of templates in 99 (GS) and 98 (raw) of the 100 cases, diverging
only when the penalty is weak. For comparison, on the same linear basis, the
reference \texttt{ElasticNetCV} selects 15--17\% of templates over the 100 realizations, and the IFS
criterion of WH21 70.5--78\%. This limitation of the IFS estimator was in fact
noted by WH21 themselves, and its origin discussed further in
Appendix~\ref{ap:whifs}.

The selection performance for the linear multiplicative configuration is shown
in Figure~\ref{fig:selectionrates}. For the strong $5\,\sigclean$ injections, both hybrid
methods select the injected templates in 97--100\% of cases, and select
uninjected templates at the $\sim 3.5\%$ level (for the GS kit) as established by their
calibration. \texttt{ElasticNetCV} selects the injected templates in 100/100
cases, but also selects uninjected templates at a rate of 40--70\%, notably
higher than the false-positive rate for the same estimator as assessed from the
uncontaminated realizations. The WH21 IFS criterion also selects 100/100
injected templates, at the cost of 57--93\% selection rate for uninjected
templates. At the weaker $2\,\sigclean$ injection, the hybrid estimators still
recover the dust extinction template in 89--100\% of cases, but select the depth proxy in
only 59--68\% of cases, and the stellar density template in 18--25\%. This
reveals the cost of the significantly decreased false-positive rate, in the form
of a loss of sensitivity to weaker contaminants (though we again emphasize that
this tradeoff may now be calibrated, rather than practitioners being at the
mercy of the intrinsic properties of the estimators). On the raw kit, the EN
hybrid selector co-selects HI4PI and extinction in 16\% of realizations at $\xi=12$,
a consequence of the high correlation $\rho = 0.92$ of the dust extinction and
neutral hydrogen maps.

The linear coefficient recovery is shown in Figure~\ref{fig:coeff-recovery}. For the strong
$5\,\sigclean$ injections, the hybrids recover the coefficients with mean ratios 1.00
to 1.01 to the input for the GS kit. \texttt{ElasticNetCV} suffers
from a coefficient shrinkage penalty, recovering a mean of 0.90 to 0.98 (this shrinkage
being one of the initial motivators of the OLS re-fit step). This bias is smaller than the noise of a single realization, but compared with an OLS re-fit on the same realizations, \texttt{ElasticNetCV} under-recovers the stellar coefficient by $0.110 \pm 0.006$, leaving ${\sim}11\%$ of the contaminant amplitude. The WH21 IFS
criterion is unbiased (ratios 1.00 to 1.01). At $2\,\sigclean$, the hybrids'
unconditional stellar density mean recovery falls to 0.3--0.4, tending to
recover either 1.6 to 1.7 times the input coefficient, or a zeroed coefficient. The
coefficients are more aggressively pulled towards zero, leaving a skewed
distribution of surviving nonzero coefficients, as intended.

Figure~\ref{fig:tradeoff} examines the tradeoffs between false-positive rate and
selection as a function of the penalty scale. As was the case for the selection
at the default scale, dust extinction is the most robust to overly-aggressive
penalties, being selected in $\geq 99\%$ of cases for the Gram-Schmidt kit, for
both injection amplitudes. Stellar contamination at $2\,\sigclean$ falls from a
selection rate of 0.84 at $\xi = 2$ (versus 0.25 clean) to 0.25 at $\xi = 12$
and 0.03 at $\xi = 24$, and similar for the depth proxy. At $5\,\sigclean$, it
remains around 0.98 through $\xi = 12$, falling to 0.72 by $\xi = 32$. In this
plane, \texttt{ElasticNetCV} lies above the hybrid curves, selecting stellar
density in 85\% of realizations in the weak case (0.16 in the clean case)
while the hybrids reach about 70\%. The IFS selector on the raw kit chooses the
weak extinction injection less often than the EN selector, though this is
reversed for the weak stellar and depth-proxy injections. This is reflective of
the difference of the two selectors in the cases of correlated template bases,
and the diagnostic power of this difference is one of the reasons for examining
both. The fact that the hybrid estimators converge for the GS kit suggests that
the orthogonalization is behaving as intended.

\begin{figure*}
  \includegraphics[width=\textwidth]{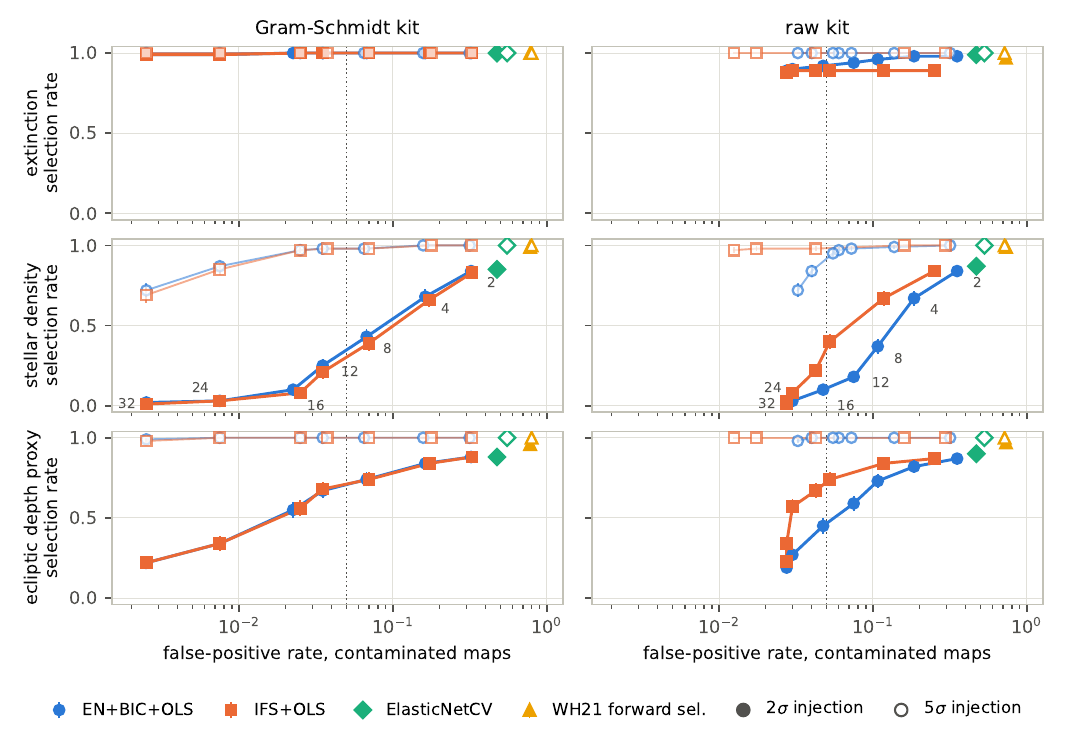}
  \caption{\small Selection rate plotted as a function of contaminated-mock false-positive
    rate, for the strong (faded, open-symbol) and weak (solid, closed-symbol)
    injections, for both kits and all three linearly injected templates. The
    EN+BIC+OLS and IFS+OLS hybrid estimators are again shown as functions of
    $\xi$ in blue and
    orange-red, respectively, as are \texttt{ElasticNetCV} in green and WH21 IFS
  in yellow-orange. $\pm1\sigma$ error bars over the 100 realizations are
  plotted but are generally small with respect to the values.
  The dotted black vertical line represents the 5\% fiducial threshold. The
  particular differences from Figure~\ref{fig:tradeoff} are a function of the
  template kit, but we emphasize the nature of the qualitative
  differences.}
  \alttext{A 3-row by 2-column matrix of plots of injected-template selection rate versus contaminated-mock false positive rate as a function of penalty $\xi$. The rows show, from the top, the extinction, stellar density, and ecliptic depth proxy contaminants, while the left column shows the Gram-Schmidt orthogonalized template kit and the right the raw kit. Extinction is detected robustly by all estimators at all penalty scales in the GS kit, with the calibrated selectors EN+BIC+OLS (blue) and IFS+OLS (orange) being slightly below 100\% recovery for the raw kit. The strong injections are recovered well at all $\xi$ for the depth proxy for both calibrated estimators, but dip slightly at higher $\xi$ for the stellar density contaminant. For both the weak stellar density and weak ecliptic depth proxy cases for both kits, 
  the trend is a general slope upwards of increasing false-positive rate with increasing selection efficiency, eventually matching the \texttt{ElasticNetCV} (green) in selection while not being able to achieve as high a false-positive rate, nor does either match the recovery or false-positive rate of WH21 IFS (yellow) which is again far to the top-right. On the raw kit for the weak injections, the false-positive rate now bottoms-out around $3\%$, and the IFS+OLS outperforms EN+BIC+OLS notably; this is resolved and the estimators again aligned and able to achieve much lower false-positive rates, extending far to the left, for the GS orthogonalized case.}
  \label{fig:tradeoff-contamfp}
\end{figure*}

Figure~\ref{fig:tradeoff-contamfp} presents a similar view as
Figure~\ref{fig:tradeoff}, but where the false-positive rates at a given $\xi$
are instead assessed against the values for uninjected templates on contaminated
mocks. On the GS kit at the same calibration $\xi=12$, the hybrid methods select uninjected
templates in 3.5\% of contaminated realizations, compared to 3.8\% from clean
mocks, and the correspondence holds at every scale. \texttt{ElasticNetCV}, on
the other hand, sees a clear degradation in its false-positive performance in
the presence of other contamination in the map, with its false positive rate
increasing from 0.15 on clean mocks to 0.48 (0.55) at 2 (5) $\sigclean$. In this
regime, the more realistic one, the hybrid estimators outperform the literature
estimators in this plane. At the most permissive scale tested, $\xi=2$, the
hybrids selected uninjected templates at a rate of 0.32 against 0.48 for
\texttt{ElasticNetCV}, with a similar 0.84--0.85 recovery of the weakly injected
stellar contaminant. WH21 IFS maintained a false-positive rate of 0.7--0.8 with
a high recovery rate, a function of the stopping criterion rather than the
contamination. On the raw kit, the false-positive rate stops falling with $\xi$,
hitting a floor. This floor is driven by HI4PI, selected in 11--14 per cent of
realizations at these $\xi$ through its correlation with the injected extinction template
($\rho = 0.92$), and depends on the selector. The orthogonalization accounts for this, when no penalty can.

\subsection{Automatic determination of non-linear order}\label{sec:res-non-lineardet}

\begin{figure}
   \includegraphics[width=\columnwidth]{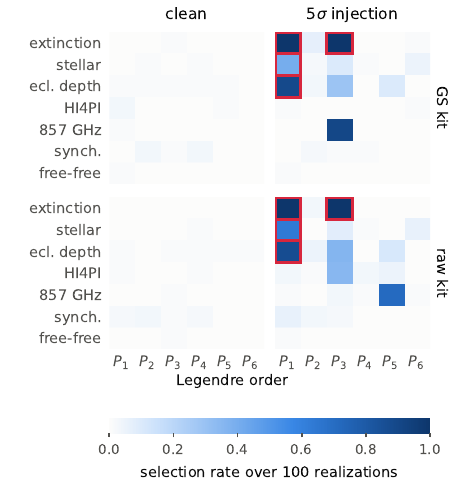}
  \caption{\small Automated order selection. Selection rates for the EN+BIC+OLS method
    for the suite of 7 templates and the six Legendre modes $P_k$. Contamination
    is injected linearly in $P_1$ for extinction, stellar
    density, and the depth proxy, and a third-order pattern $P_3$ is also injected for
    extinction (injected terms shown as red square outlines). The left column
    shows the null on clean mocks. The non-linear extinction terms are recovered
    robustly, though errant higher-order terms are often selected in
    place of the stellar density $P_1$ term. For the \textit{Planck} 857 GHz
    template, this false-positive rate has a physical origin due to its
    relationship with the extinction template, discussed further in Section~\ref{sec:res-templatecorr}.}
    \alttext{A $2\times2$ matrix of polynomial order $\times$ template sub-matrices, with the left column being the clean and the right column the $5\sigclean$ injections, and the top row the GS kit with the bottom row the raw kit. The sub-matrices show each of the seven templates, the first three contaminating, on the vertical axis, and the polynomial order $P_1$--$P_6$ on the horizontal axis. Each cell of the sub-matrices is shaded blue based on the selection rate over the 100 mocks. Injected orders are highlighted in red. Very little signal is seen in the clean-mock column. For the injected column, the injected coefficients are typically detected robustly, except for the $P_1$ component of stellar contamination, and there is also clear shading from uninjected templates, particularly \textit{Planck} 857 GHz.}
  \label{fig:ordersel}
\end{figure}

Figure~\ref{fig:ordersel} shows the per-order selection on the non-linear contaminated
set, where the extinction component is injected with both a first- and a
third-order density modulation, using the same rescaled spatial template. The extinction
response is injected directly in the fitting basis, so this is a test of how the basis
expansion composes with the selection for known Legendre-order couplings, rather than a
statement about the actual physical response to dust extinction. For
the $5\,\sigclean$ injection, both $P_1$ and $P_3$ injected components of dust
extinction were correctly selected in 100/100 mocks, for both the GS and raw
template bases, with uninjected orders for those templates selected at a level consistent with the
scatter in the mocks. The injected linear component of stellar contamination is only
recovered in 40\% (GS) and 63\% (raw) of mocks, a result of the linear-injected
stellar power entering the $P_3$ term.

\textit{Planck} 857 is selected in 90/100 cases at $P_3$ in the GS case, and in
71/100 cases at $P_5$ in the raw case. A clear extension to guard against
this could take the form of applying the orthogonalizing pre-treatment to the
full $K=6$ Legendre basis. The non-linear basis expansion and automated selection
are working together to inform the relationships between the templates in
our kits.

\begin{figure*}
  \includegraphics[width=\textwidth]{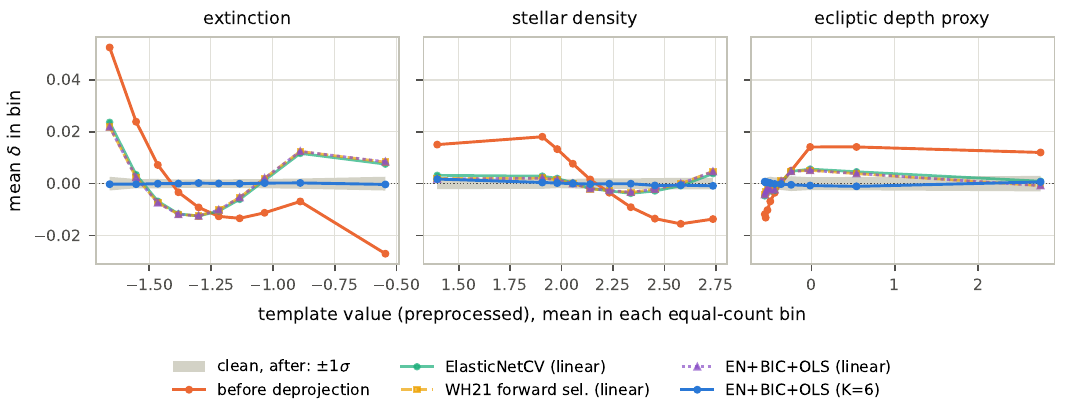}
  \caption{\small Overdensity binned in equal-pixel-count bins of each systematic
    template's values, before and after deprojection with the four methods
    (\texttt{ElasticNetCV} in green, WH21 IFS in yellow-orange, EN+BIC+OLS with
    the linear basis in purple, and EN+BIC+OLS with the K=6 Legendre basis in
    blue, shown as the means over 100 mock realizations) for deprojection with
    the GS kit. Extinction, stellar density, and the ecliptic depth proxy are
    injected at $5\,\sigclean$. The clean-mock $\pm 1\sigma$ distribution is
    shown as the grey shaded region. The linear methods choose the same basis on
  average for the $5\,\sigclean$ injection, so their results are similar.}
  \alttext{3 figures side-by-side showing the overdensity on the vertical axis binned as a function of each systematic's value shown on the horizontal axis, with extinction, stellar density, and the ecliptic depth proxy running from left to right. Each figure has an orange curve showing the density modulation due to the systematic before deprojection. A green solid line shows the modulation after the application of the \texttt{ElasticNetCV} estimator, a yellow dashed line shows the same for the WH21 forward selection estimator, and a purple dotted line shows that for the EN+BIC+OLS estimator, all using a linear basis. A blue solid line shows EN+BIC+OLS on the Legendre $K=6$ basis, and a grey shaded band shows the $\pm1\sigma$ distribution of the 100 mock realizations. The pre-cleaning density variations are far beyond the mocks in all cases. The linear-basis methods all perform visually nearly identically, and are within a few $\sigma$ of the mocks in a few bins for stellar density and the depth proxy, but leave significant remaining signal in the extinction overdensity profile. The blue curve denoting the non-linear method sits well inside the mock distribution at all values, for all templates.}
  \label{fig:binneddensity} 
\end{figure*}

Figure~\ref{fig:binneddensity} shows the corrected map-level contrast statistic at the
fiducial $\xi = 24$ penalty for the $5\,\sigclean$ injections, using the
EN+BIC selector with the full $K=6$ GS basis. In all cases the residuals are
appropriately corrected to a degree consistent with the clean mocks' scatter $\sigma$, with the
maximum post-deprojection bin fluctuating at $0.84\,\sigma$ (stellar density,
lowest-value bin), compared to $8.98\,\sigma$ for the pure-linear
\texttt{ElasticNetCV}, $8.30\,\sigma$ for the WH21 IFS, and $8.31\,\sigma$
for linear EN+BIC+OLS (all for extinction). The difference is principally the
basis rather than the selector/estimator, but we
highlight the aggressive automated curation of the hybrid approaches as enabling
this flexible approach.

\begin{figure*}
  \includegraphics[width=\textwidth]{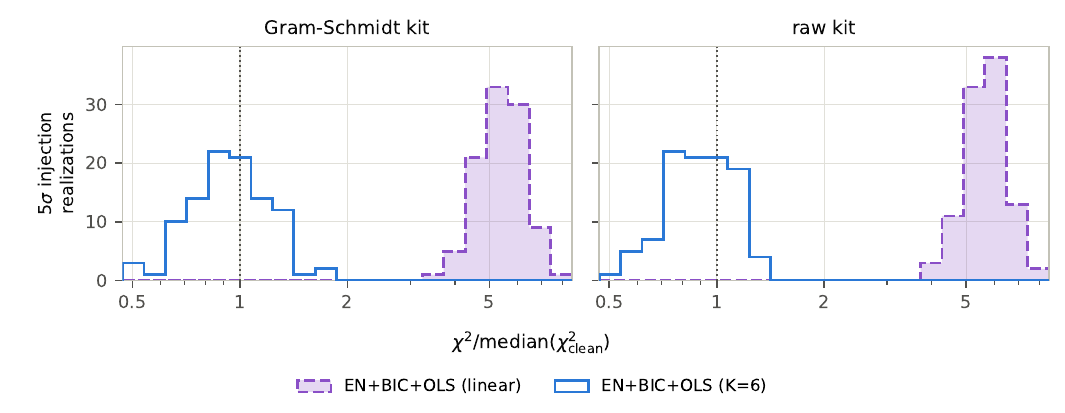}
  \caption{\small Mahalanobis $\chi^2$ ratio of the post-cleaned residual maps over the
    template set compared to the uncontaminated mocks deprojected similarly.
    The EN+BIC+OLS hybrid is shown with its linear basis in purple and its
    K=6 Legendre basis in blue. With the purely linear basis, the non-linear
    injected coupling is not
    able to be fit, biasing the linear basis high. The uncorrected median values
  are centred at 14.1 (14.0) for the $5\,\sigclean$ injection for the GS (raw) case.}
  \alttext{Two side-by-side histograms of Mahalanobis $\chi^2$ ratios comparing cleaned $5\sigclean$-contaminated realizations to uncontaminated mocks for the Gram-Schmidt kit (left) and the raw kit (right). The EN+BIC+OLS is applied with a linear basis, shown as a purple shaded histogram, and with the $K=6$ Legendre basis, shown as an unfilled blue histogram. The non-linear method is centred on and clustered around 1 for both cases, while the linear method's distribution is centred many times higher, near or beyond 5.}
  \label{fig:recovery}
\end{figure*}

Figure~\ref{fig:recovery} shows the Mahalanobis $\chi^2$ of the same binned map
statistics, comparing the linear ($\xi=12$) with the non-linear ($K=6,\,\xi=24$) EN+BIC+OLS
deprojection approaches. The (artificial, $P_1 + P_3$) non-linear coupling of the dust
extinction pattern is
unable to be cleaned by the linear basis, with $\chi^2 / {\rm median}(\chi^2_{\rm clean})$ ratios compared to
the identically deprojected clean mocks of 5.33 (5.69) for the $5\,\sigclean$
contaminants for the GS (raw) kit. The
flexible non-linear basis, on the other hand, captures this readily, with
$\chi^2 / {\rm median}(\chi^2_{\rm clean})$ ratios of 0.94 (0.90) for the
$5\,\sigclean$ injections for the GS (raw) kits. The non-linear basis leaves
2--7\% of realizations beyond the clean-mock 95th percentile, while every linear
realization is beyond the highest-fluctuating clean mock.

\subsection{Step-edge structure and the B-spline basis}
\label{sec:res-non-linear}

\begin{figure*}
  \includegraphics[width=\textwidth]{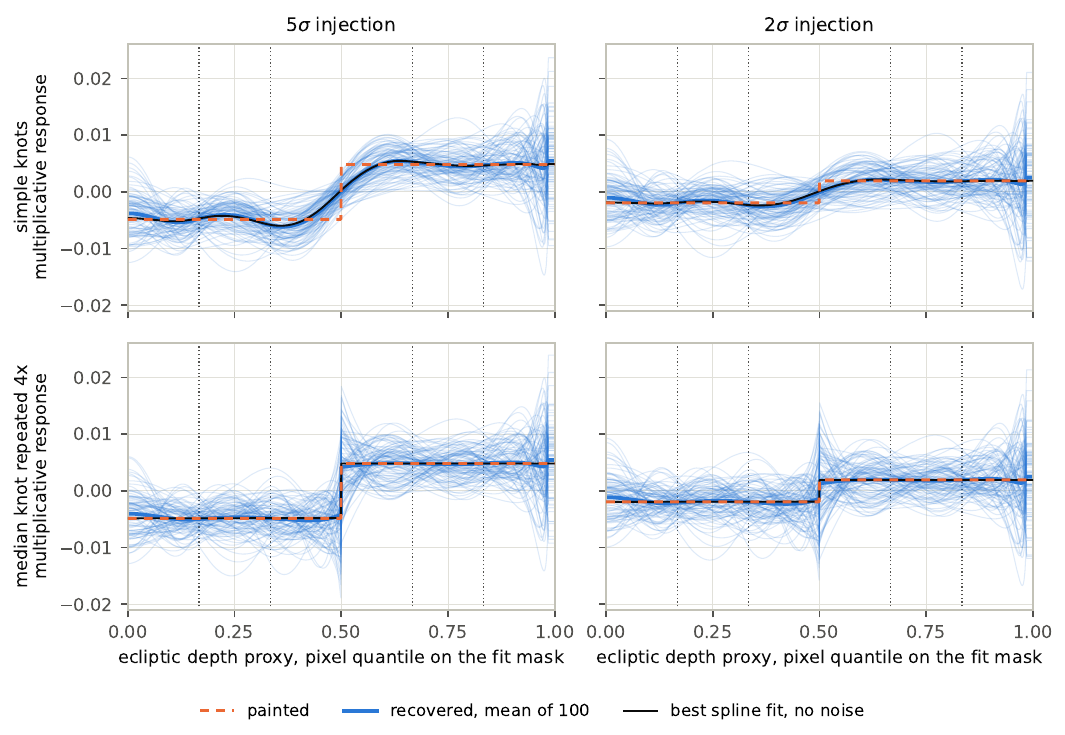}
  \caption{\small A multiplicative response injected as a step function in the ecliptic
  depth quantile, shown in orange, deprojected with a cubic B-spline
  with individual realizations shown in light blue and the mean in
solid blue. Vertical dotted lines denote knot locations. The top row shows a simple five-knot configuration, while the bottom
row adds additional knots at the location of the step to better fit the
sharp feature. The ideal spline fit in the presence of no mock signal is shown as a
dotted black line, generally lying underneath the mock mean. The step is
recovered robustly in the mean for the added-knot case, with some clear spline
overshoot for the simple-knot case.
}
\alttext{A 2$\times$2 array of subplots, each showing the overdensity response in the presence of a response with a step at the template's median and an associated B-spline fit. The left column shows the strong $5\sigclean$ injections and the right the weak $2\sigclean$, and the top row shows a simple 5-knot cubic spline configuration while the bottom row shows a version adding additional knots at the location of the step. An orange dashed line shows the injected step, a black line the ideal best spline-fit without any noise or cosmology, thin blue lines individual realizations' fits, and a thick blue line the mean of the 100 realizations. The simple-knot configuration captures the step approximately with some error around the step itself, while the repeated-knot configuration captures it sharply in both cases. The step is much smaller in the $2\sigclean$ case, but the result is similar.}
  \label{fig:stepedge}
\end{figure*}

Figure~\ref{fig:stepedge} shows the step-edge contaminant OLS fit with a cubic
B-spline, which for the standard configuration uses 5 knots, placing the central
one at the centre of the step. A more complicated knot configuration replicates
the central knot 4 times (3 additional times; \emph{i.e.}, 8 knots total) to better capture the sharp transition. This is
indicative of a known edge in response to a template, such as a discontinuity
due to a sample cut in an imaging survey, for which
an analyst may have some prior knowledge of where to place the knots. The basis
can represent the step to an RMS error over pixels relative to the step height
of 0.236 (5 knots) or 0.021 (8 knots)  without noise, and the mean over realizations reaches 0.24 (0.27) for
the $5\,\sigclean$ ($2\,\sigclean$) cases, for the simple knot
configuration. The added-knot configuration reaches 0.06 (0.14) for the
$5\,\sigclean$ ($2\,\sigclean$) cases. Individual realizations
are noisier, with median errors over individual realizations of 0.59 (1.39) and
0.61 (1.52) for the $5\,\sigclean$ ($2\,\sigclean$) injections for the simple
and added-knot cases, respectively.

In this case, the step is a property of the bulk behaviour of the knots below the
edge and those above it, with each feature only carrying a share of the
response. If these are offered as separate
components to the automated fitting selection, only one or a
couple of columns tend to be selected at reasonable $\xi$. Since the placement
of knots is best informed manually, anyway, they may similarly be passed
directly to the regression step, bypassing the selector, to avoid this.

\subsection{Higher-order template-template coupling}
\label{sec:res-templatecorr}

\begin{figure*}
  \includegraphics[width=\textwidth]{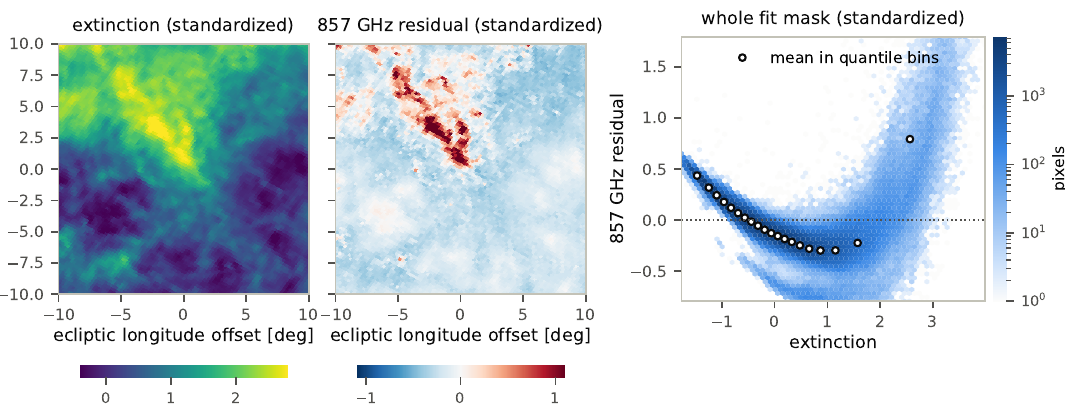}
  \caption{\small Higher-order template coupling. The left panel shows the
    (primary, never orthogonalized) CSFD dust extinction map and the central
    panel the post-orthogonalized residual \textit{Planck} 857 GHz map. The former
  traces dust extinction and the latter emission, so they may be anticorrelated
  in diffuse (low-column-density) regions and strongly correlated in dense
  regions, which is made clear by plotting them against one another as in the
  right-hand panel showing a 2-D hexagonal histogram of pixel counts and means
  in equal-pixel quantile bins. The Gram-Schmidt orthogonalization procedure may remove the
linear cross-correlation, but higher-order spatial relationships may persist.}
\alttext{Three panels showing, from the left, zoom-ins of the CSFD dust extinction and orthogonalized \textit{Planck} 857 GHz map, and a plot of the two's values against each other as a 2-D pixel-binned histogram with denser regions shaded darker blue. Bright, dense clumps in the two map zoom-ins are positively signed in both maps, while the diffuse structure is positive in the CSFD map and negative in the \textit{Planck} map. This results in a U-shaped relationship in the right-hand panel, with the orthogonalized residual of the \textit{Planck} map rising sharply to positive values at very low and very high values of the dust extinction, with a less-extreme negative minimum in intermediate values.}
  \label{fig:extinction857}
\end{figure*}

The curation of Section~\ref{sec:curation} orthogonalizes the
\textit{Planck} 857 GHz map against dust extinction, stellar density, the depth
proxy, and HI4PI, with a residual fraction of 0.82. The orthogonalized remainder
of the template has negligible correlation with extinction by construction
($r^2 \sim 0$ to machine precision), and a na\"{\i}ve linear curation would identify the two as
independent. Figure~\ref{fig:extinction857} shows that this is not the full story. Plotting
the orthogonalized \textit{Planck} 857 template values against the extinction
template values reveals a clear parabolic relationship in extinction, with
positive \textit{Planck} map values at both the lowest and highest extinction
values, but negative between. The 857 GHz map traces dust emission, which scales
with column density and a temperature-dependent emissivity, while the CSFD
extinction map traces reddening, which scales with the column density alone. The
diffuse, low-density regions then often end up anticorrelated in practice, while
high-density regions are highly positively correlated, explaining the structure
in the figure.

Plotting each orthogonalized residual against the values of the ``primary''
templates, and those of the primary templates against each other, is then a
useful way of screening for this type of effect. This issue can likely be readily addressed by extending the orthogonalization procedure along the $K$ direction, which we leave to future work.

\subsection{Additive and multiplicative contamination}
\label{sec:res-addmult}

We apply the joint maximum-likelihood estimator of Section~\ref{sec:regress} to
the mock contaminated with the additive stellar component, and have it operate
on a fixed linear basis of the contaminating templates (the matched control) to
isolate the effect of the estimator. With templates matching the additive contaminant's spatial profile, the
multiplicative-only and joint fits leave residuals that agree to 1.5 per cent
of the clean scatter, but differ in the second-moment attribution of
Section~\ref{sec:fom}. The multiplicative-only fit gives a median of 0.94 (0.97) at
$2\,\sigclean$ ($5\,\sigclean$), while the joint estimator recovers the true
attribution with a median of
$-0.07$ ($-0.07$). At this injection level, the modulation is 0.7 per cent rms (3 per
cent at the extreme edge of the mask).

\subsection{Signal suppression}
\label{sec:res-transfer}

\begin{figure*}
  \includegraphics[width=\textwidth]{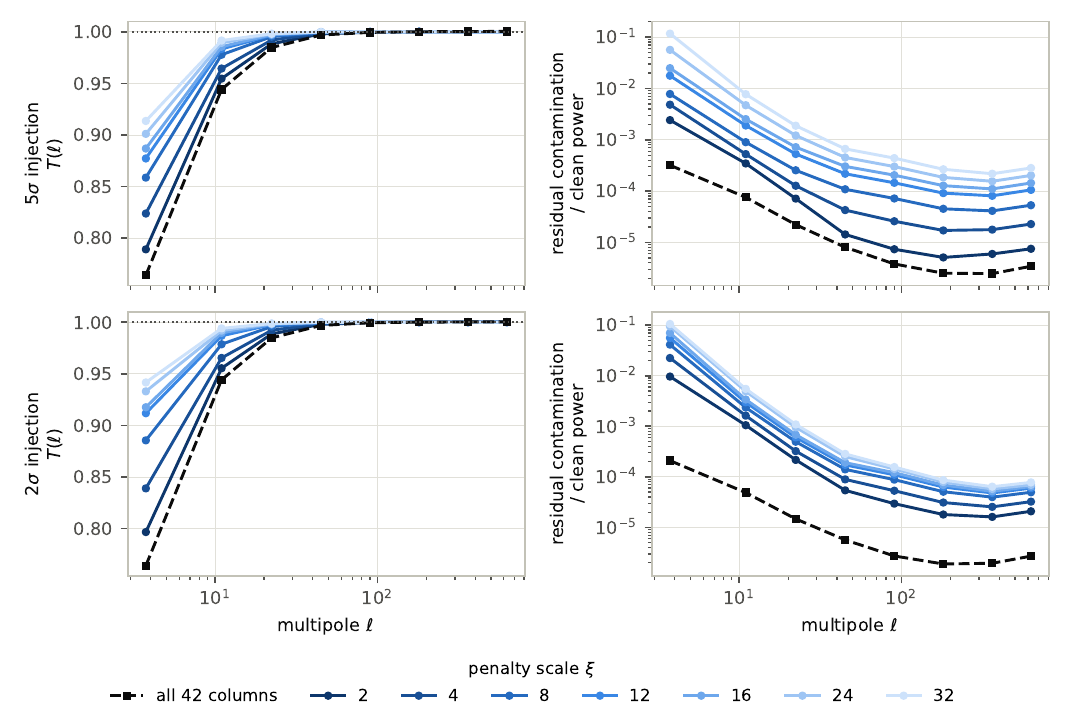}
  \caption{\small Signal transfer function and residual contamination, described in Section~\ref{sec:fom}.(iv). The left-hand
    column shows the signal transfer function, $T(\ell)$, for the EN+BIC+OLS selector on the
    $K=6$ Legendre basis on the Gram-Schmidt kit, as a function of the
    penalty scale for the $5\,\sigclean$ (top) and $2\,\sigclean$ (bottom) cases. Lines are
    shown as means over the 100 mock realizations; the dashed line being that for an OLS on all
    42 template-order components. The right-hand column shows
    the resulting residual contamination as a fraction of the clean angular
    power. The higher floor of the selected estimator in the $2\,\sigclean$ case is a result of
    unselected contaminating templates, with a false negative in 6 of
    100 realizations even at $\xi=2$.}
    \alttext{A $2\times2$ array of, on the left, signal transfer function $T(\ell)$, and on the right, residual contamination line plots as a function of angular multipole $\ell$. The top row shows those for the $5\sigclean$ injections and the bottom for the $2\sigclean$. For each, different penalty scales $\xi$ are shown with darker shades of blue denoting smaller values, and a black dashed line showing a forced OLS regression of all 42 template-order basis components. The $T(\ell)$ plots begin at around 0.8 below multipoles $\ell \sim 10$ before rising quickly, with larger values of $\xi$ levelling out at 1 more rapidly, within a few percent at $\ell=10$ for $\xi=32$. The plots look similar between the two contamination cases. The residual contamination plots are U-shaped but with a general slope downwards with increasing $\ell$, again stratified by $\xi$ but with the all-column-regression case clearly lying below.}
  \label{fig:transferfunc}
\end{figure*}

The ultimate consequences of our penalty scale tuning are illustrated in
Figure~\ref{fig:transferfunc}, showing the angular transfer functions and residual power levels of Section~\ref{sec:fom}.(iv). Template deprojection represents a balancing act
between preserving input cosmological power, and correctly removing the
contamination due to the systematics being deprojected. At $5\,\sigclean$, 21.1\%
of power in the lowest ($\ell = 2$--7) multipole bin is removed at $\xi=2$ and 8.6\% at
$\xi=32$, quickly falling towards higher multipoles. For the low
signal-to-noise $2\,\sigclean$ injections, fewer templates pass the selection
stage, and the loss is 20.3 to 5.8\% over $\xi = 2$--32 in the lowest-multipole
bin. Deprojecting all 42 template-order components removes 23.6\% of power in this bin. Residual contamination in that same lowest-$\ell$
bin rises from $2.4 \times 10^{-3}$ of
the clean power at $\xi=2$ to $1.2 \times 10^{-1}$ at $\xi=32$ for the
$5\,\sigclean$ injection, and $9.6 \times 10^{-3}$ to $1.05 \times 10^{-1}$ for the $2\,\sigclean$ case,
compared with $3.2 \times 10^{-4}$ ($5\sigclean$) and $2.1 \times 10^{-4}$ ($2\sigclean$) for the 
42-component case. Over the range in
$\xi$ considered from 2--32, the signal cost in the largest bin changes by about
12.5 percentage points, while the residual changes by a factor of 50. Note that the precise nature
of these curves depends not just on the scaling with $\xi$
but on the patterns of the initial set of templates offered,
so we emphasize that this is intended to be schematic and should typically be
assessed on an analysis-by-analysis basis for the particular ensemble under consideration.

\section{Discussion and conclusions}
\label{sec:conclusions}

Separating the selection of templates from the deprojection step allows the
calibration of the false-positive rate using a single mock-calibrated scale, and
provides unbiased contaminant recovery. At the fiducial 5\% false-positive
calibration target, the two information-criterion penalized hybrid methods
presented here select uninjected templates 4 to 5 times less often than
\texttt{ElasticNetCV} on clean maps and more than 10 times less often on
contaminated maps, while recovering $5\,\sigclean$ contaminants in 97 to 100\% of
realizations. The price paid for this is the low selection rate of some low signal-to-noise templates, a
tradeoff illustrated in Figure~\ref{fig:tradeoff}.

The two hybrid selectors approach selection differently, using
information-criterion penalized versions of the ElasticNet and IFS algorithms,
while then employing the same cleaning step (OLS or the maximum-likelihood
estimator). For well-calibrated selectors, these selections should converge,
providing an additional consistency check and informing template kit curation.
They might be expected to diverge in the case of highly correlated input
template sets, and we indeed find that agreement between the two selectors is
visibly aided by applying a Gram-Schmidt orthogonalization procedure to the
template set as a preprocessing step.

Combining the automated down-selection with a Legendre expansion of the template
basis allows the data to set the required non-linear order, while still limiting
the overall number of degrees of freedom to a reasonable number and providing
outputs that are usefully interpretable. The retained order may capture a
variety of non-linear relationships of the density coupling to the systematic
contaminants, as well as absorbing the impact of transforms or preprocessing
steps that might push that relationship from linearity. A B-spline basis option
enables straightforward handling of features like sharp discontinuities in
imaging depth, though the knots require manual placement.

The orthogonalization of the template basis can aid in recovering
weak injections and can stabilize the estimators, though a simple linear orthogonalization may neglect more
complicated template-template relationships arising from real astrophysics or
other effects, and so may be readily composed with the non-linear expansion.

With this particular template set, the suppression of power with more aggressive
selection is typically confined to $\ell < 32$, suppressing the mock power by
about 8.6--21.1\% (5.8--20.3\%) over $\xi =$2--32 for $5\,\sigclean$ ($2\,\sigclean$) contaminants
in the largest-scale bin, compared to 23.6\% (both cases) when deprojecting all 42 templates.
The level of contamination may change
by orders of magnitude as various templates are or
are not selected. For a realistic survey, the power suppression must be assessed on a case-by-case basis on simulations for a production selected template basis.

\textbf{Limitations.} The levels of contamination, significant in a
$\delta{n}/\bar{n}$ sense even for $\sim1\,\sigclean$ for many systematics, and
the relative insensitivity of map-based regression approaches below
$\sim2\,\sigclean$ highlight the limitations of this data-driven approach. Some
templates like dust extinction are highly relevant but nearly always selected; some are marginally selected but have a somewhat less significant impact. The
truly dangerous class both modulate the density field significantly on large
scales and may be missed some fraction of the time. Analyst judgement is still useful in electing to force the
selection of certain templates that are expected to impact the data but for
which detection is uncertain, and in practice it is likely prudent to set the penalty scale $\xi$ rather conservatively to prioritize suppressing contamination. Fitting across additional redshift bins would
likely aid this, as would approaches that marginalize over some reasonable set
of priors for the template ensemble coupling amplitudes. Additional null and
internal consistency testing of the eventual data vectors should be essential
to establish the robustness of and to unblind cosmological results on data.

The mocks are simple lognormal mock catalogues with linear
input power spectra and no dynamics. The template kit contains seven plausible
templates, but is by no means comprehensive, and the particular scale
dependences and residuals are highly interrelated with the composition of this
initial kit. Contamination here is purely angular, without consideration of a
radial component, which is relevant for SPHEREx and will be the subject of
dedicated future work. The $\chi^2$ statistic here is first-moment in the
templates.

As mentioned in Section~\ref{sec:contam}, even a weak level of contamination may approach the systematics budgets of next-generation efforts towards $\fnl$, with $1\sigclean$ accounting for SPHEREx's entire $\delta n / \bar{n}$ budget in the particularly extreme case of dust extinction. A variety of approaches, templates, and consistency tests must ultimately be employed as part of the strategy to control potential biases from systematic effects. A generally under-accounted-for caveat to the template deprojection
procedure as typically laid out is the lack of consideration for error in the
fitted values (\emph{i.e.}, in the templates themselves) and how this may bias
the regression, as well as the impact of similar effects from templates which
represent decorrelated proxies of the true patterns of density modulations. To meet the requirements for SPHEREx's target of $\sigma(\fnl)<0.5$, we anticipate
this to be another direction of significant future effort.

\section*{Acknowledgements}

The authors wish to thank Henry Gebhardt, Olivier Dor\'e, and Yun-Ting Cheng for useful feedback on the manuscript. The authors acknowledge support from the SPHEREx project under a contract from the
NASA/Goddard Space Flight Center to the California Institute of Technology.
JRC wishes to acknowledge support from the David \& Ellen Lee Fellowship. CWS
acknowledges support from the David and Lucile Packard Foundation, and NASA grant
22--ROMAN11--0011, via a subaward from the Jet Propulsion Laboratory. The
High Performance Computing resources used in this investigation were provided
by funding from the JPL Information and Technology Solutions Directorate.
Computations were run on the \textsc{Frontera}, \textsc{Stampede3}, \textsc{Lonestar6}, and
\textsc{Vista} clusters provided by the Texas Advanced Computing Center
(TACC)\footnote{\url{http://www.tacc.utexas.edu/}}. The authors would
like to acknowledge the use of
Anthropic's\footnote{\url{https://www.anthropic.com/}} \textsc{Claude Opus} 4.8, 5.0,
and \textsc{Fable} 5 models for assistance in code implementation, refinement of
figures, and review of the manuscript. All such output was reviewed by the authors who take full responsibility for the content.
  
\section*{Code and Data Availability}

The SPHEREx template deprojection pipeline is implemented within the SPHEREx Level 4
cosmology pipeline and is not currently made public, though an attempt has been
made to describe the algorithm in sufficient detail to facilitate
reimplementation. The results shown here are
purely algorithmic on mocks with no SPHEREx data used. Mock catalogs and derived
products can be made available upon reasonable request to the corresponding author.

\bibliographystyle{mnras}
{\interlinepenalty=10000
\bibliography{deproj_main}}

%%%%%%%%%%%%%%%%%%%%%%%%%%%%%%%%%%%%%%%%%%%%%%%%%%

%%%%%%%%%%%%%%%%% APPENDICES %%%%%%%%%%%%%%%%%%%%%

\appendix
\section{The forward-selection criterion of WH21}
\label{ap:whifs}

WH21 perform a linear fit to the unbinned contrast against each template, and
using a Wald statistic for a parameter $\theta$,
$\Delta \chi^2 = \mathbf{\theta}^{\rm T} \mathbf{C}_{\theta}^{-1}\mathbf{\theta}$,
to score the slope and intercept against zero, select the largest, reweight, and
iterate until all satisfy $\Delta \chi^2 / \Delta \chi^{2}_{0} < 2$, where
$\Delta \chi^2_0$ is the 68th percentile of $\Delta \chi^2$ in the absence of
contamination. The ``DES Y1'' method of WH21 uses clean mocks to assess this,
while WH21's iterative forward selection instead takes $\mathbf{C}_\theta$
directly from the fit itself, treating pixels as independent with equal variance
and estimating $\chi^2_0$ from the null distribution analytically.

If the contrast is properly zero-mean by construction, then this has only one degree of
freedom, the slope. In the case of uncorrelated pixels, for which the
assumptions are valid, $\Delta \chi^{2}_{0} = 0.989$ (the 68th percentile of a
$\chi^2$ distribution with one degree of freedom), the threshold is 1.98, and
the per-template false-positive rate at the published threshold is 16\%. Because
cosmic variance is absent from the null, in practice the criterion
selects most templates on all of the clean mocks (Figure~\ref{fig:methodcomp}).
WH21 anticipate exactly this effect in their Appendix~B, and discuss
prewhitening steps that may be taken to address this, noting that the similar DES Y1
method shown therein already incorporates a normalisation from uncontaminated
mocks.

%%%%%%%%%%%%%%%%%%%%%%%%%%%%%%%%%%%%%%%%%%%%%%%%%%

% Don't change these lines
\bsp	% typesetting comment
\label{lastpage}
\end{document}